\documentclass[conference]{IEEEtran}
\IEEEoverridecommandlockouts
\usepackage{cite}
\usepackage{amsmath,amssymb,amsfonts}
\usepackage{algorithmic}
\usepackage{textcomp}
\usepackage{xcolor}
\usepackage[pdftex]{graphicx}
\usepackage{epstopdf}

\usepackage{amsmath,amssymb,enumerate} 
\usepackage{bm} 
\usepackage{indentfirst} 
\usepackage{empheq} 
\usepackage{bm}
\usepackage{comment}
\usepackage{here} 
\newcommand{\dd}{\mathrm{d}}
\newcommand{\e}{\mathrm{e}}
\newcommand{\ii}{\mathrm{i}}

\newcommand{\R}{\mathbb{R}}
\newcommand{\C}{\mathbb{C}}

\def\BibTeX{{\rm B\kern-.05em{\sc i\kern-.025em b}\kern-.08em
    T\kern-.1667em\lower.7ex\hbox{E}\kern-.125emX}}
\begin{document}

\title{Increase of Low-Frequency Modes of User Dynamics in Online Social Networks During Overheating of Discussions
}

\author{\IEEEauthorblockN{Masaki Aida}
    \IEEEauthorblockA{
        \textit{Tokyo Metropolitan University} \\
Tokyo 191--0065, Japan \\
aida@tmu.ac.jp}
\and
\IEEEauthorblockN{Koichi Nagatani}
    \IEEEauthorblockA{
        \textit{Tokyo Metropolitan University} \\
Tokyo 191--0065, Japan \\
ksts0826lv@gmail.com}
\and
\IEEEauthorblockN{Chisa Takano}
    \IEEEauthorblockA{
        \textit{Hiroshima City University} \\
Hiroshima 731-3194, Japan \\
takano@hiroshima-cu.ac.jp}
}

\maketitle

\begin{abstract}
User dynamics in online social networks have a significant impact on not only the online community but also real-world activities.
As examples, we can mention explosive user dynamics triggered by social polarization, echo chamber phenomena, fake news, etc. 
Explosive user dynamics are frequently called online flaming.
The wave equation-based model for online social networks (called the oscillation model) is a theoretical model proposed to describe user dynamics in online social networks.
This model can be used to understand the relationship between explosive user dynamics and the structure of social networks.
However, since the oscillation model was introduced as a purely theoretical model of social networks, it is necessary to confirm whether the model describes real phenomena correctly or not.
In this paper, we first show a prediction from the oscillation model; the low-frequency oscillation mode of user dynamics will be dominant when the structure of online social networks changes so that user activity is activated.
To verify the predictions with actual data, we show spectral analyses of both the log data of posts on an electronic bulletin board site and the frequency data of word search from Google Trends. 
The results support the predictions from the theoretical model. 
\end{abstract}


\section{Introduction}
\label{sec:introduction}
As information networks penetrating everyday life, it has dramatically increased the exchange of information among individuals, and the user dynamics of online social networks (OSNs) are beginning to have a major impact not only on online communities but also on social activities in the real world. 
In particular, explosive user dynamics such as online flaming is often seen in OSNs and can be grown much faster than human rational decision making can respond, which can cause major social unrest. 

Here, online flaming means the phenomena that the activity of user dynamics becomes extremely high due to the impact of user networking. 
Therefore understanding the user dynamics of OSNs based on a mathematical model is an urgent and important issue.
This paper focuses on the behaviors of user dynamics. 

Online flaming is generally taken to be a sudden flood of hostile messages targeting the same victim. 
To fully understand flaming, it may be necessary to consider the background of the incident from the perspective of sociology and behavioral psychology, because flaming is caused by people's thoughts and activities. 
Unfortunately,  this consideration is just a post-incident coping approach. 
To establish a timely countermeasure to flaming, a mathematical model that can describe flaming without analyzing the behavior of people is necessary. 

Let us consider the question of whether we can take measures against flaming without analyzing the behavior of people. 
As an example, consider the bandwidth design of the classical telephone network. 
Although telephone calls are generated from the activities of individual human beings, we have successfully modeled them as random stochastic processes without modeling human behavior. 
If this example is followed, it may be possible to understand online flaming in some kind of mathematical framework rather than by understanding human behavior. 
If successful, it will be the basis for developing timely countermeasures to flaming. 

Online user dynamics, including explosive user dynamics, are generated by interactions between users. 
It is difficult to fully understand the details of interactions between actual users, but we can apply the concept of the minimal model; it models the simple interactions commonly included in many diverse types of user interactions. 
The minimal model was used to develop the oscillation model on networks to describe online user dynamics \cite{aida2018}. 
In the oscillation model approach, network dynamics is described by the wave equation on networks. 

\begin{figure}[bt]
\begin{center}
\includegraphics[width=0.8\linewidth]{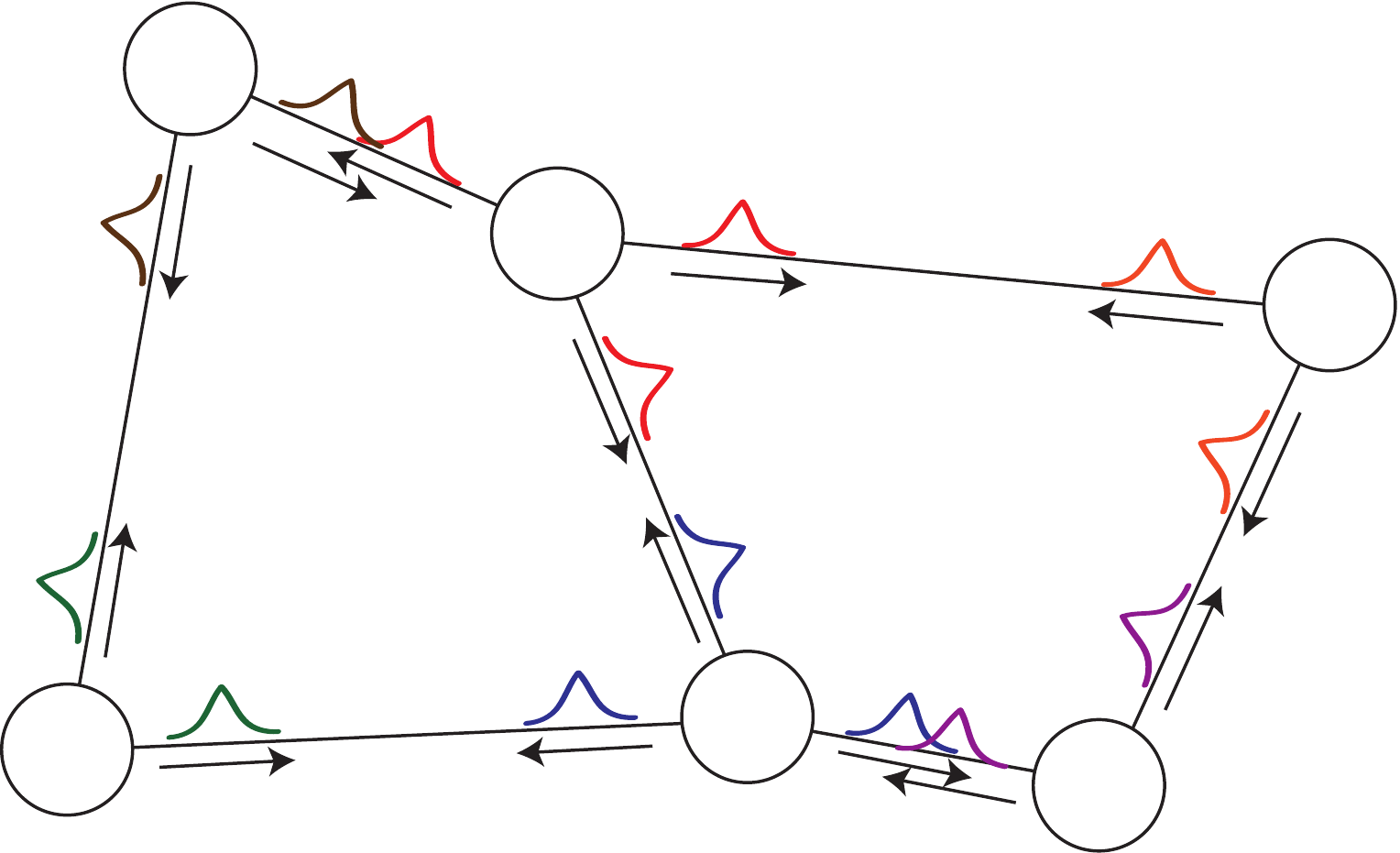}
\caption{Propagation of influence between users at finite speeds (node: user, link: the relationship between users)}
\label{app:oscillation_model}
\end{center}
\end{figure}

The reason why the wave equation on networks is applied in the oscillation model describing online user dynamics is as follows. 
OSN users interact and have some influence on each other.
Such influences should be propagated through the relationship of users, in other words, through OSNs. 
Figure~\ref{app:oscillation_model} illusrates such propagations of influences. 
The propagation of influence never occurs at an infinite speed; it would occur at finite speeds. 

\begin{figure}[bt]
\begin{center}
\includegraphics[width=0.8\linewidth]{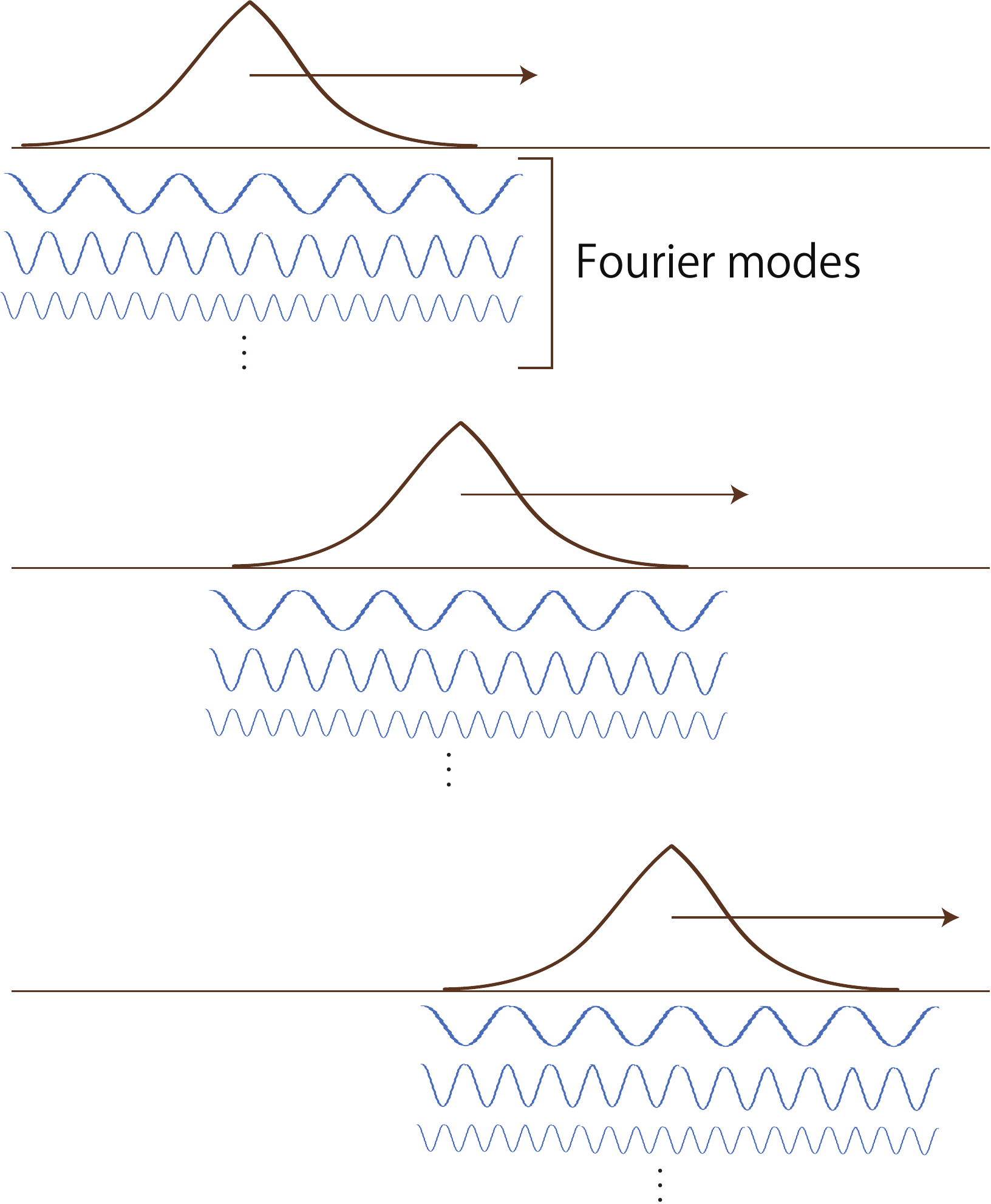}
\caption{Propagation and the wave equation}
\label{app:wave_eq}
\end{center}
\end{figure}

In general, the wave equation describes the propagation of something through a medium. 
In our case, it is the influence of users propagated through OSNs.
Figure~\ref{app:wave_eq} shows the situation that some influence of a user propagates at finite speeds from left to right. 
Since the waveform of influence can be decomposed into Fourier modes, each Fourier mode should propagate at finite speeds if the influence propagates at finite speeds. 

There is significant a difference between the oscillation model for OSNs and the wave equation on networked objects in the real world. 
Since the strength of influence between users is asymmetry, Newton's third law about action and reaction is violated in OSNs in general, unlike physical models in the real world. 
This causes the violation of the energy conservation law in OSNs and causes online flaming as a result. 

The validity of the oscillation model has some theoretical support because it provides a natural derivation of conventional node centrality \cite{globecom2016,takano2018}. 
Node centrality shows the importance of individual nodes and many node centrality measures have been proposed in response to the sheer variety of definitions of node importance. 
Conventional examples are the degree centrality and the betweenness centrality \cite{centrality1,centrality2,wasserman,carrington,mislove}. 
The oscillation energy of each node calculated from the oscillation model gives both the conventional degree centrality and the betweenness centrality in special cases.
In addition, the oscillation energy gives a generalization of node centrality. 
In some conditions, the oscillation model can describe the characteristic of oscillation energy divergence in certain OSN configurations. 
This means the emergence of explosive user dynamics can be explained by OSN structure \cite{aida2018,aida-book2020,aida2017}.  

As mentioned above, the oscillation model is simple but its theoretical framework has the potential to be able to describe explosive user dynamics.
However, the oscillation model has yet to be validated by using actual data to confirm that it can describe real phenomena correctly. 
In this paper, we first derive a prediction from the oscillation model; the low-frequency oscillation mode of user dynamics will be dominant when the sudden increase of user activity in OSN is triggered.
To verify the predictions against actual data, we conduct spectral analyses of both the log data of posts on an electronic bulletin board site and the frequency data of word search from Google Trends. This paper is an extended version of the research presented in \cite{nagatani,nagatani_bigdata_2019}. 
The previous studies introduced some experimental results but there were some problems; the reason behind the selection of the experimental data was unclear, and there were no comparisons with control groups in which user dynamics were not triggered. This paper provides experimental results to rectify the omissions of those papers.

The rest of this paper is organized as follows. 
Sect.~\ref{sec:RW} shows related work to clarify the position of the oscillation model. 
Sect.~\ref{sec:OM} briefly summarizes the oscillation model. 
Sect.~\ref{sec:lf} shows the prediction that the low-frequency mode of user dynamics becomes prominent based on the oscillation model. 
Sect.~\ref{sec:exprmnt} shows experimental results and clarifies that they do not contradict the prediction of the oscillation model. 
Finally, we conclude discussions in Sect.~\ref{sec:conclusion}. 

\section{Related Work}
\label{sec:RW}
Spectral graph theory is one of the key approaches for investigating the structure of networks and/or dynamics on networks, and the eigenvalues and the eigenvectors of the Laplacian matrix play important roles in investigating network structure and/or dynamics. 
One of the most significant properties of spectral graph theory is the fact that we can introduce graph Fourier transformation \cite{Hammond}, which is the diagonalization of the Laplacian matrix. 

Graph Fourier transformation has accelerated the development of graph signal processing \cite{Shuman,Sand2013,Sand2014}, which is a generalized signal processing for graph-structured data. 
Graph signal processing can generate a transformation that yields both multi-resolution views of graphs and an associated multi-resolution of a graph signal \cite{Luxburg,Bruna,Shuman2016}. 
This is applicable to multi-layer clustering for social networks. 

Recently, graph neural networks (GNNs) have been actively developed \cite{Zhang}. 
This is a combination of the graph signal processing and deep learning technology. 
In particular, graph convolutional networks (GCNs) are a significant topic in GNNs.
First introduced in \cite{Bruna}, convolution for graph signals is based on graph Fourier transformation. 
As an application of GNNs to OSNs, \cite{Fan2019} discusses a mechanism for social recommendations. 

In this paper, we are interested not in deep learning for graph data but in user dynamics in OSNs, that is, the dynamic aspect of user activities in OSNs.    

Studies on user dynamics in OSNs have examined various models in an effort to capture the diversity of the characteristics of user dynamics. 
User dynamics that describe the adoption and abandonment of a particular social networking service (SNS) have been modeled by the SIR model, which is a traditional epidemiological model, and the irSIR model, which is an extension of that model~\cite{SIR,Nekovee2007,Cannarella2014EpidemiologicalMO}. 
These models express the state transition of an objective system in a macroscopic framework, but are not good at describing individual user dynamics. 
Also, these models describe the speed of changes in transient states and/or the configuration of the final steady-state of the system, but they fail to address the divergence of user dynamics. 

The consensus problem including user opinion formation is typical of the dynamics in OSNs~\cite{Olfati-Saber2004,Wang2010}. 
This can be modeled by a first-order differential equation with respect to time by using a Laplacian matrix that represents the social network structure.
The differential equation used in this model is a sort of a continuous-time Markov chain on the network.
First-order differential equations with respect to time are also used in modeling of the temporal change in social network structure (linking or delinking of the nodes), and there are models that consider the change via a continuous-time Markov chain~\cite{Snijders2010}. 
In addition to theoretical models, \cite{Cha2009} and \cite{Zhao2012} studied user dynamics analysis based on real network observations. 
Similar to epidemiological models, the Markov chain describes the transient states and/or the steady-state of the system, but not the divergence of user dynamics. 

The significant advantage of the oscillation model based on the wave equation in OSNs is that it can describe the process by which the strength of user dynamics diverges.

\section{Oscillation Model of Online User Dynamics}
\label{sec:OM}
The user dynamics in OSNs are complex, and understanding them completely is as difficult as understanding human thoughts and human interactions. 
For this reason, we give upset aside tackling the complete description of user dynamics and aim at the simplest model possible that can describe the properties that are common to many types of user interaction.
This approach is called a {\it minimal model}. 
As a minimal model, the oscillation model was proposed to describe user dynamics in OSNs and takes the form of a wave equation on networks. 

\subsection{Wave Equation of the Oscillation Model}
According to \cite{aida2018}, an outline of the oscillation model is summarized as follows. 

\begin{figure}[b]
\begin{center}
\includegraphics[width=0.70\linewidth]{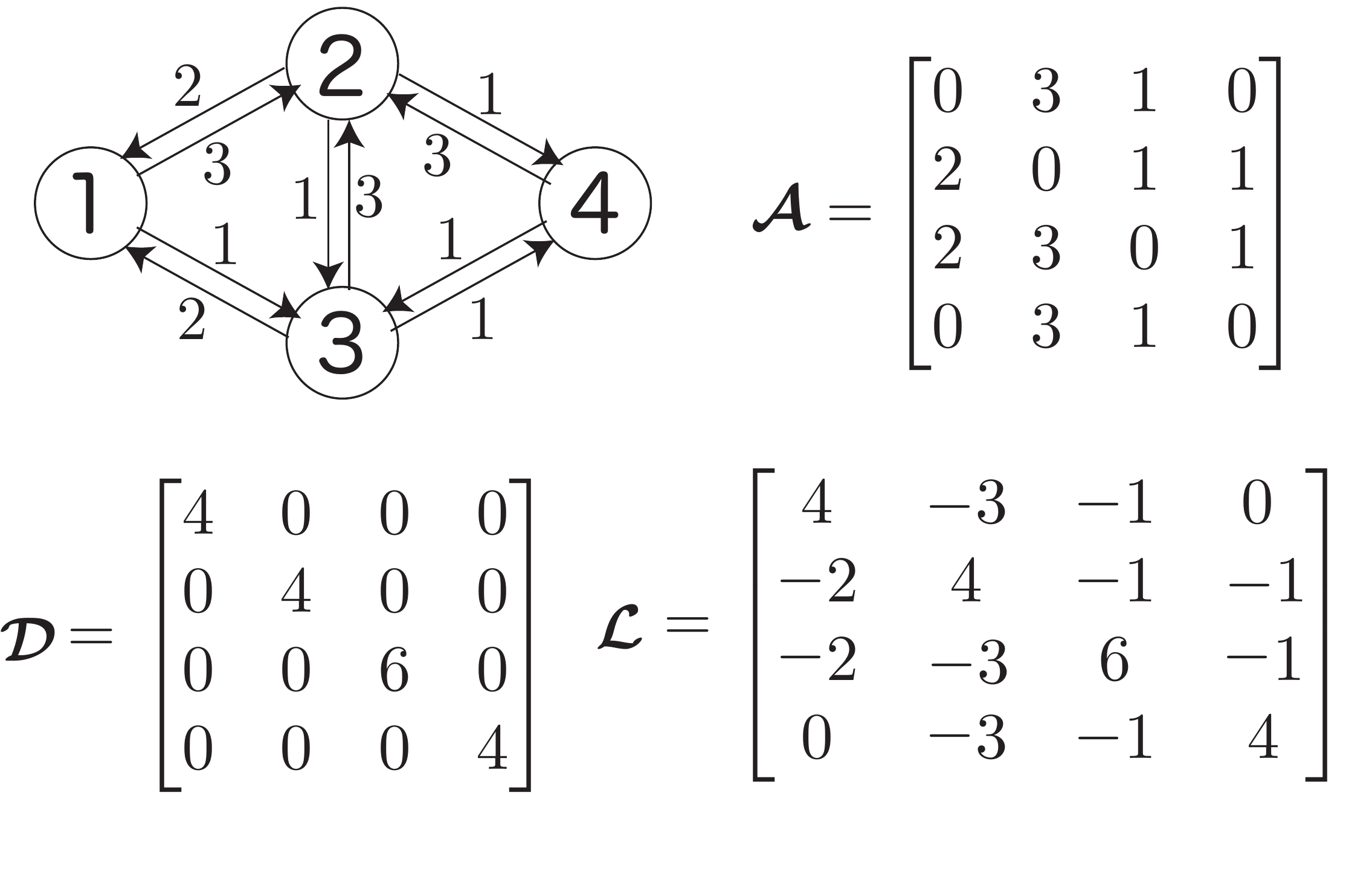}
\caption{An example of the Laplacian matrix}
\label{fig:ex}
\end{center}
\end{figure}

Let an OSN be modeled by a directed graph $G(V, E)$ with $n$ nodes, where $V=\{1,\,\dots,\,n\}$ is the set of nodes and $E$ is the set of directed links. 
For nodes $i,\,j \in V$, the weight of the directed link $(i \rightarrow j) \in E$ is $w_ {ij}$ $(>0)$, 
The adjacency matrix $\bm{\mathcal{A}} = [\mathcal{A}_{ij}]_{1\le i,j \le n}$ of the OSN is defined as 
\begin{align}
\mathcal{A}_{ij} := \left\{
\begin{array}{cl}
w_{ij},&  \quad (i\rightarrow j) \in E,\\
0,& \quad (i\rightarrow j) \not\in E.
\end{array}
\right. 
\end{align}
Also, for the nodal (weighted) out-degree $d_i := \sum_{j\in \partial i} w_{ij}$, 
the degree matrix is defined as $\bm{\mathcal{D}} := \mathrm{diag}(d_1,\,\dots\,d_n)$. 
Here, $\partial i$ denotes the set of adjacent nodes of out-links from node $i$. 
Then, the Laplacian matrix of the directed graph representing the structure of the OSN is defined 
by $\bm{\mathcal{L}} := \bm{\mathcal{D}} - \bm{\mathcal{A}}$. 
An example of the Laplacian matrix is shown in Fig.~\ref{fig:ex}. 

Let the state vector of users at time $t$ be $\bm{x}(t) := {}^t\!(x_1(t),\,\dots,\,x_n(t))$, where 
$x_i(t)$ $(i=1,\,\dots,\,n)$ denotes the user state of node $i$ at time $t$. 
Then, the wave equation on the OSN is written as 
\begin{align}
\frac{\dd^2}{\dd t^2} \, \bm{x}(t) = - \bm{\mathcal{L}} \, \bm{x}(t). 
\label{eq:wave-eq}
\end{align}
In general, the wave equation describes the propagation of something through a medium at finite speeds. 
In this case, the wave equation (\ref{eq:wave-eq}) describes that the influence between users propagates through OSN at finite speeds. 
Since no influence can propagate at an infinite speed in general, the wave equation-based model is a natural way of thinking. 

We can give another reason for the use of a wave equation.
Let us consider user state $x_i(t)$ for node $i$, and $x_j(t)$ for its adjacent node $j$. 
We assume the interaction between the adjacent nodes act so that the difference $\Delta_{ij} := x_i(t) - x_j(t)$ changes toward $0$. 
In addition, for the force acting at node $i$ from node $j$, $F_{ij}$, we assume the strength of $F_{ij}$ is a function of the difference $\Delta_{ij}$, and $F_{ij} = 0$ for $\Delta_{ij} = 0$. 
The Taylor expansion of $F_{ij}$ is given as 
 \begin{align}
F_{ij}(\Delta_{ij}) &= -w_{ij} \, \Delta_{ij} + o(\Delta_{ij}) 
\notag\\
&\simeq -w_{ij} \, (x_i(t) - x_j(t)), \quad (\Delta_{ij}\ll 1),
\label{eq:taylor}
\end{align}
where $w_{ij} > 0$. 
Although $F_{ij}(\Delta_{ij})$ is a nonlinear function of $\Delta_{ij}$, the linear approximation (\ref{eq:taylor}) is valid at least for small $\Delta_{ij}$. 
Therefore, the dynamics obtained from (\ref{eq:taylor}) is commonly applicable to many types of user interaction. 
The equation of motion of node $i$ is written as 
 \begin{align}
\frac{\dd^2}{\dd t^2} \, x_i(t) &= \sum_{j\in\partial i} F_{ij}(\Delta_{ij}) 
\notag\\
&= -\left(d_i \, x_i(t) - \sum_{j\in\partial i} w_{ij} x_j(t)\right). 
\end{align}
This equation is equivalent to (\ref{eq:wave-eq}). 

\subsection{Oscillation Energy and Node Centrality}
Although the oscillation model is simple, it has been confirmed that it has the following advantages.
\begin{enumerate}[$\bullet$]
\item The oscillation energy for each node calculated from the oscillation model not only provides a common framework for the conventional indices of node centrality (degree centrality and betweenness centrality \cite{wasserman,carrington,mislove}) but also can provide an extended concept of node centrality that is applicable to OSNs with more complex configurations \cite{globecom2016,takano2018}. 
\item it is possible to describe the phenomenon in which the activity of user dynamics grows explosively as seen in the online flaming phenomenon in OSNs~\cite{aida2018}.
\end{enumerate}

To explain the above advantages briefly, we first define here the symmetrizable directed graph. 
Since the row sum of the Laplacian matrix $\bm{\mathcal{L}}$ is zero, one of the eigenvalues of $\bm{\mathcal{L}}$ is $0$. 
Thus, we let ${}^t\!\bm{m} = (m_1,\,\dots,\,m_n)$ be the left eigenvector associated with the eigenvalue of 0, that is, 
\[
{}^t\!\bm{m} \, \bm{\mathcal{L}} = (0,\,\dots,\,0). 
\]
The directed graph is symmetrizable if and only if there is an $\bm{m}$ all of whose components are positive $m_i > 0$ $(i = 1,\,\dots,\,n)$ and satisfy 
\[
m_i \, w_{ij} = m_j \, w_{ji},
\]
for all adjacent nodes $i$ and $j$. 
Hereafter, the Laplacian matrix for a symmetrizable directed graph is denoted as $\bm{\mathcal{L}}_0$. 

\begin{figure}[tb]
\begin{center}
\includegraphics[width=1.00\linewidth]{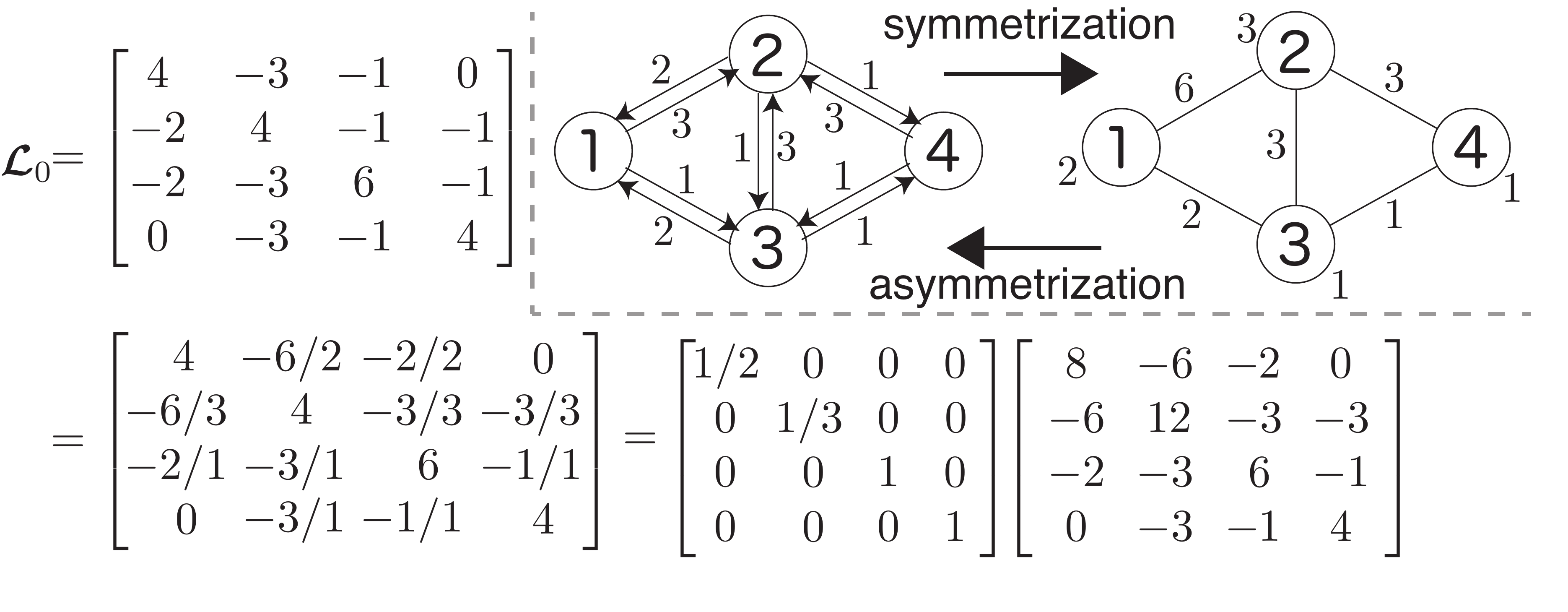}
\caption{An example of the symmetrization of the Laplacian matrix for a directed graph.}
\label{fig:ex2}
\end{center}
\end{figure}

Let $\bm{M}$ be a diagonal matrix, $\bm{M} := \text{diag}(m_1,\,\dots,\,m_n)$, whose diagonal components are the components of $\bm{m}$. 
The Laplacian matrix $\bm{\mathcal{L}}_0$ for any symmetrizable directed graph can be written as the product of $\bm{M}^{-1}$ and a (symmetric) Laplacian matrix $\bm{L}$ for a certain  undirected graph, as follows;  
\begin{align}
\bm{\mathcal{L}}_0 = \bm{M}^{-1} \, \bm{L}. 
\label{eq:M^{-1}L}
\end{align}
An example of this decomposition is shown in Fig.~\ref{fig:ex2}. 

\begin{figure}[tb]
\begin{center}
\includegraphics[width=0.8\linewidth]{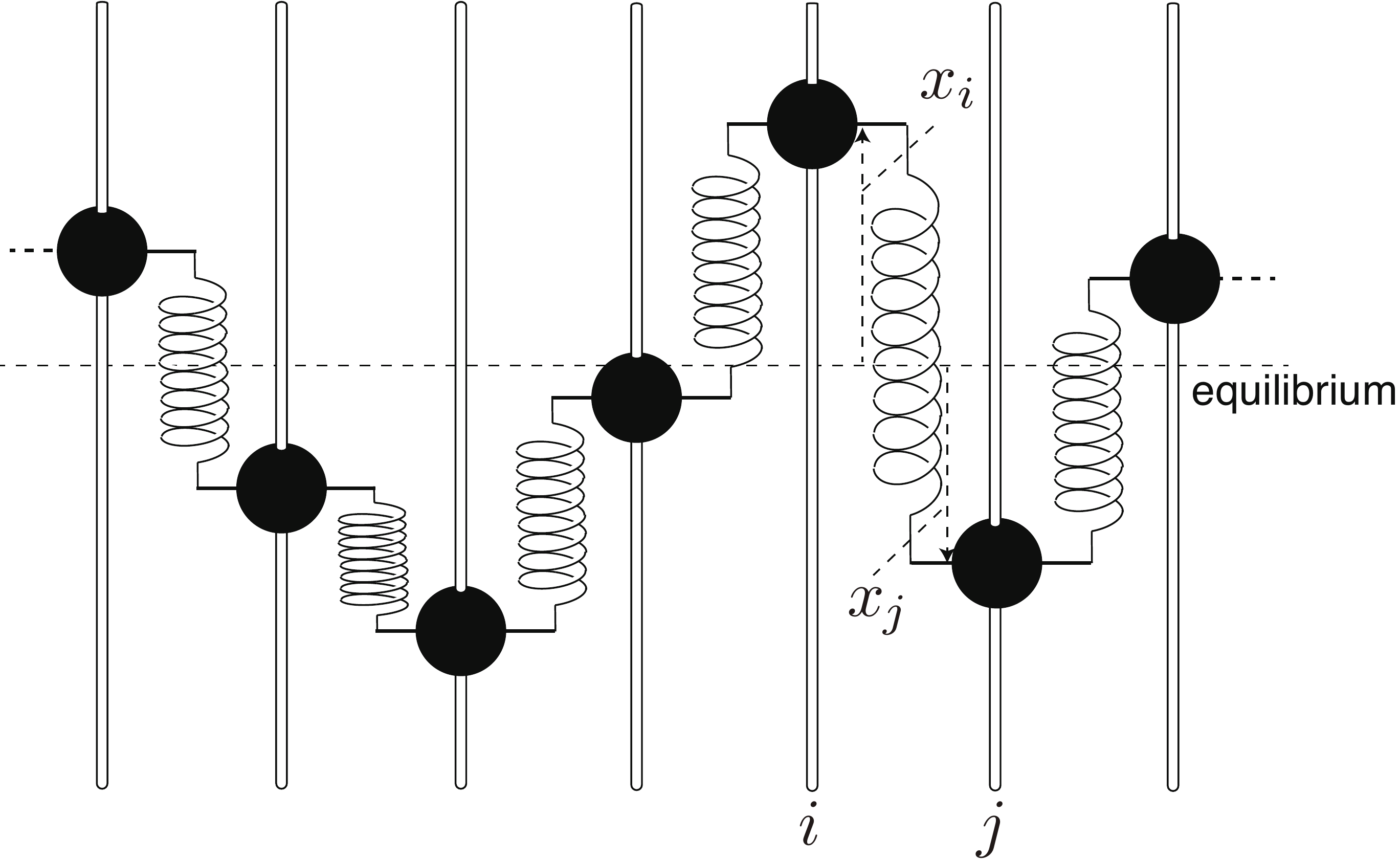}
\caption{Simple example for 1-dimensional network.}
\label{fig:oscillator}
\end{center}
\end{figure}

The wave equation (\ref{eq:wave-eq}) for symmetrizable directed graph $\bm{\mathcal{L}}_0$ can be written as
\begin{align}
\frac{\dd^2}{\dd t^2} \, \bm{x}(t) = - \bm{\mathcal{L}}_0 \, \bm{x}(t). 
\label{eq:wave-eq0}
\end{align}
Rewriting the wave equation (\ref{eq:wave-eq0}) using (\ref{eq:M^{-1}L}) gives
\[
\bm{M}\,\frac{\dd^2}{\dd t^2} \, \bm{x}(t) = - \bm{L} \, \bm{x}(t). 
\]
From this, it can be seen that $\bm{M}$ corresponds to the masses of nodes and the link weights of the undirected graph represented by $\bm{L}$ corresponds to the spring coefficients. 
Figure~\ref{fig:oscillator} illustrates an example situation using a simple 1-dimensional network. 
Here, we introduce the scaled Laplacian matrix $\bm{S}_0$ for symmetrizable directed graph $\bm{\mathcal{L}}_0$ as
\begin{align}
\bm{S}_0 := \bm{M}^{+1/2} \, \bm{\mathcal{L}}_0 \, \bm{M}^{-1/2}, 
\label{eq:S_0}
\end{align}
where $\bm{M}^{+1/2} := \text{diag}(\sqrt{m_1},\,\dots,\,\sqrt{m_n})$. 
For symmetrizable directed graph $\bm{\mathcal{L}}_0$, scaled Laplacian matrix $\bm{S}_0$ is a real symmetric matrix. 
The wave equation (\ref{eq:wave-eq0}) can, by using $\bm{S}_0$, be written as 
\begin{align}
\frac{\dd^2}{\dd t^2} \, \bm{y}(t) = - \bm{S}_0 \, \bm{y}(t),  
\label{eq:wave-eq-s}
\end{align}
where $\bm{y}(t) := \bm{M}^{+1/2} \, \bm{x}(t)$. 

To solve the wave equation (\ref{eq:wave-eq0}), we solve (\ref{eq:wave-eq-s}) first. 
Let us define the eigenvalues and eigenvectors of $\bm{S}_0$. 
Since $\bm{S}_0$ is a real symmetric matrix, all the eigenvalues are real numbers. 
In addition, the minimum value of the eigenvalue is $0$. 
Thus we sort the eigenvalues as 
\[
0 = \lambda_0 \le \lambda_1 \le \cdots \le \lambda_{n-1}, 
\]
where the multiplicity of eigenvalues of $0$ is the number of connected components of the OSN. 
Let the eigenvector associated with eigenvalue $\lambda_\mu$ be $\bm{v}_\mu$. 
Since $\bm{S}_0$ is a real symmetric matrix, we can choose eigenvectors $\{\bm{v}_\mu\}_{\mu=0,\,1,\,\dots,\,n-1}$ as the orthonormal eigenbasis. 
That is, eigenvectors are mutually orthogonalized and their lengths are $1$;  
\[
\bm{v}_\mu \cdot \bm{v}_\nu = \delta_{\mu\nu}. 
\]
$\bm{S}_0$ is obtained by the similarity transformation (\ref{eq:S_0}) of $\bm{\mathcal{L}}_0$. 
It follows that the eigenvalues of $\bm{S}_0$ are the same as those of $\bm{\mathcal{L}}_0$. 

The solution of (\ref{eq:wave-eq-s}) can be expanded by the eigenbasis to yield 
\[
\bm{y}(t) = \sum_{\mu=0}^{n-1} a_\mu(t) \, \bm{v}_\mu. 
\]
By substituting this into (\ref{eq:wave-eq-s}), we have the following $n$ independent equations 
\begin{align}
\frac{\dd^2}{\dd t^2} \, a_\mu(t) = - \lambda_\mu \, a_\mu(t),  
\label{eq:harmonic}
\end{align}
for the oscillation modes $\mu=0,\,1,\,\dots,\,n-1$. 
Each of these equations is the equation of the motion of a simple harmonic oscillator. 
The solution of (\ref{eq:harmonic}) is obtained as 
\begin{align}
a_\mu(t) = c_\mu^+ \, \exp(+\ii\omega_\mu t) + c_\mu^- \, \exp(-\ii\omega_\mu t),  
\label{eq:sol-mode}
\end{align}
where $\omega_\mu := \sqrt{\lambda_\mu}$ is the eigenfrequency, and $c_\mu^+$ and $c_\mu^-$ are constants. 
The solution of (\ref{eq:wave-eq0}) is obtained as  
\begin{align}
\bm{x}(t) = \sum_{\mu=0}^{n-1} \left[c_\mu^+ \, \exp(+\ii\omega_\mu t) + c_\mu^- \, \exp(-\ii\omega_\mu t)\right] \bm{M}^{-1/2}\,\bm{v}_\mu.  
\label{eq:sol}
\end{align}
 
From the solution (\ref{eq:sol}), the oscillation energy $E_i$ of node $i$ for symmetrizable directed graphs is obtained as 
\begin{align}
E_i = \sum_{\mu=0}^{n-1} \lambda_\mu \left(|c_\mu^+|^2 + |c_\mu^-|^2 \right)  v_\mu(i)^2, 
\label{eq:E_i}
\end{align}
where $v_\mu(i)$ is the $i$-th component of the eigenvector $\bm{v}_\mu = {}^t\!(v_\mu(1),\,\dots,\,v_\mu(n))$. 
The oscillation energy for each node gives the conventional node centrality (degree centrality and betweenness centrality) as special cases. 
Consider the situation in which all nodes equally contribute to the initial condition, that is, any node can, with the same probability, become the source node of activity. 
If all the link weights of $\bm{\mathcal{L}}_0$ are $1$, the oscillation energy $E_i$ is proportional to the degree centrality, that is, the nodal degree $d_i$. 
If each link weight is determined as the number of shortest paths, between all pairs of nodes, that are passing through the link, the oscillation energy $E_i$ gives the betweenness centrality and a constant. 
This means the oscillation model gives a common framework for the conventional degree centrality and betweenness centrality. 
Also, as the oscillation energy can be calculated even in general complicated usage conditions, it gives a generalized notion of node centrality.

In general, the Laplacian matrix $\bm{\mathcal{L}}$ of OSNs is not necessarily a symmetrizable directed graph. 
In such situations, the wave equation on network does not correspond to physical dynamics like shown as Fig.~\ref{fig:oscillator}. 
Interaction between adjacent nodes violates Newton's third law and the energy conservation law of the system does not hold, in general. 

\begin{figure}[bt]
\begin{center}
\includegraphics[width=0.55\linewidth]{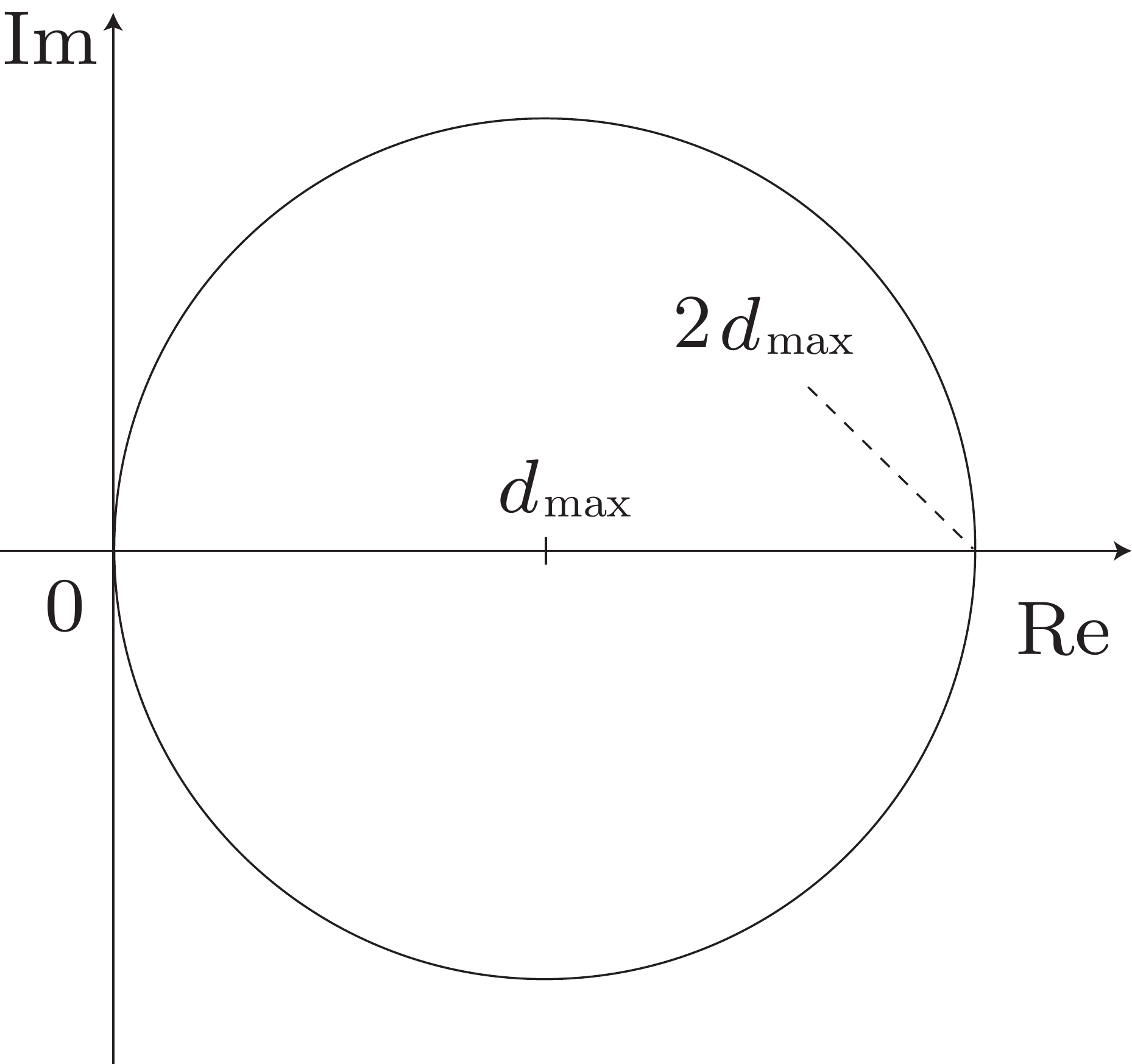}
\caption{The maximum Gershgorin disk of $\bm{\mathcal{L}}$}
\label{app:gershgorin}
\end{center}
\end{figure}

For a general directed graph, the eigenvalue of the Laplacian matrix is not restricted to real numbers. 
Even in that case, the real part of the eigenvalues is known to be non-negative.
From the Gershgorin circle theorem \cite{aida-book2020,Gershgorin_theorem}, all the eigenvalues are included in the largest Gershgorin disk on the complex plane.
Here, the largest Gershgorin disk of  the Laplacian matrix is a circle whose center is on the real axis at the maximum weighted nodal degree
\[
d_\mathrm{max} = \max_{i\in V} d_i,
\]
and its radius is also $d_\mathrm{max}$ (see Fig.~\ref{app:gershgorin}). 

If eigenvalue $\lambda_\mu$ of the Laplacian matrix $\bm{\mathcal{L}}$ for an OSN is not a real number, eigenfrequency $\omega_\mu$ is also a non-real number. 
Let the non-real eigenfrequency $\omega_\mu \in \C$ be 
\[
\omega_\mu = a + b\ii,
\]
where $a,\,b \in \R$ and $a \ge 0$. 
Note that $\omega_\mu^2 = \lambda_\mu$. 

In this case, the eigenvectors $\{\bm{v}_\mu\}_{\mu=0,\,1,\,\dots,\,n-1}$ associated with eigenvalue $\lambda_\mu$ are no longer mutually orthogonalized. 
However, we can assume that eigenvectors $\{\bm{v}_\mu\}_{\mu=0,\,1,\,\dots,\,n-1}$ are linearly independent. 
Thus we can solve the wave equation (\ref{eq:wave-eq}) similarly and the solution for oscillation mode $\mu$ is obtained as (\ref{eq:sol-mode}). 
As the eigenfrequency is not real, the solution for oscillation mode $\mu$ is written as 
\begin{align}
a_\mu(t) &= c_\mu^+ \, \exp(+\ii (a+b\ii) t) + c_\mu^- \, \exp(-\ii (a+b\ii) t)
\notag\\
&= c_\mu^+ \, \exp(-bt) \, \exp(+\ii at) + c_\mu^- \, \exp(+bt) \, \exp(-\ii at). 
\notag
\end{align}
This means that the amplitude of $a_\mu(t)$ diverges with time as $\exp(+|b| t)$, which causes the divergence of the oscillation energy.  
We consider that this process underlies the explosive user dynamics  generated in OSNs including online flaming. 

In the oscillation model, the oscillation energy $E$, which represents the strength of the users' activity, is an observable quantity, whereas the users' state, $\bm{x}(t)$, is not observable.
As mentioned earlier, $\bm{x}(t)$ is given as a solution to the wave equation, and the wave equation is required for describing the propagation of the influence between users at finite speeds.
This structure is similar to quantum mechanics.
Even in quantum mechanics, the wave function that represents the state of the system is a solution of a certain wave equation, that cannot be observed itself.
What can be observed is the square of the absolute value of the wave function corresponding to the oscillation energy.
In Sect.~\ref{sec:exprmnt}, we will verify the prediction of the oscillation model with experimental data, but in that case, the observable amount of the oscillation energy is the subject of analysis.

\section{Increase of Low-Frequency Mode in the Oscillation Model}
\label{sec:lf}
The oscillation model is derived from a purely theoretical framework. Although it is possible to form the basis of the conventional node centrality from the oscillation model, it is not yet clear whether it adequately describes the actual user dynamics of OSNs. 
This section explains the characteristics that increase the strength of the low-frequency mode predicted by the oscillation model in the situation that online user dynamics becomes highly active; the predicted characteristics are verified by assessing actual data. 

\subsection{Decomposition of the Laplacian Matrix}
We now consider how to decompose the Laplacian matrix of a directed graph into two parts: 
one is the Laplacian matrix $\bm{\mathcal{L}}_0$ for a symmetrizable directed graph and the other is the Laplacian matrix for a one-way link graph, $\bm{\mathcal{L}}_\mathrm{I}$, that is,  
\begin{align}
\bm{\mathcal{L}} = \bm{\mathcal{L}}_0 + \bm{\mathcal{L}}_\mathrm{I}, 
\label{eq:L-decomp}
\end{align}
where the one-way link graph has only one-way directed links between nodes.  

\begin{figure}[tb]
\begin{center}
\includegraphics[width=0.95\linewidth]{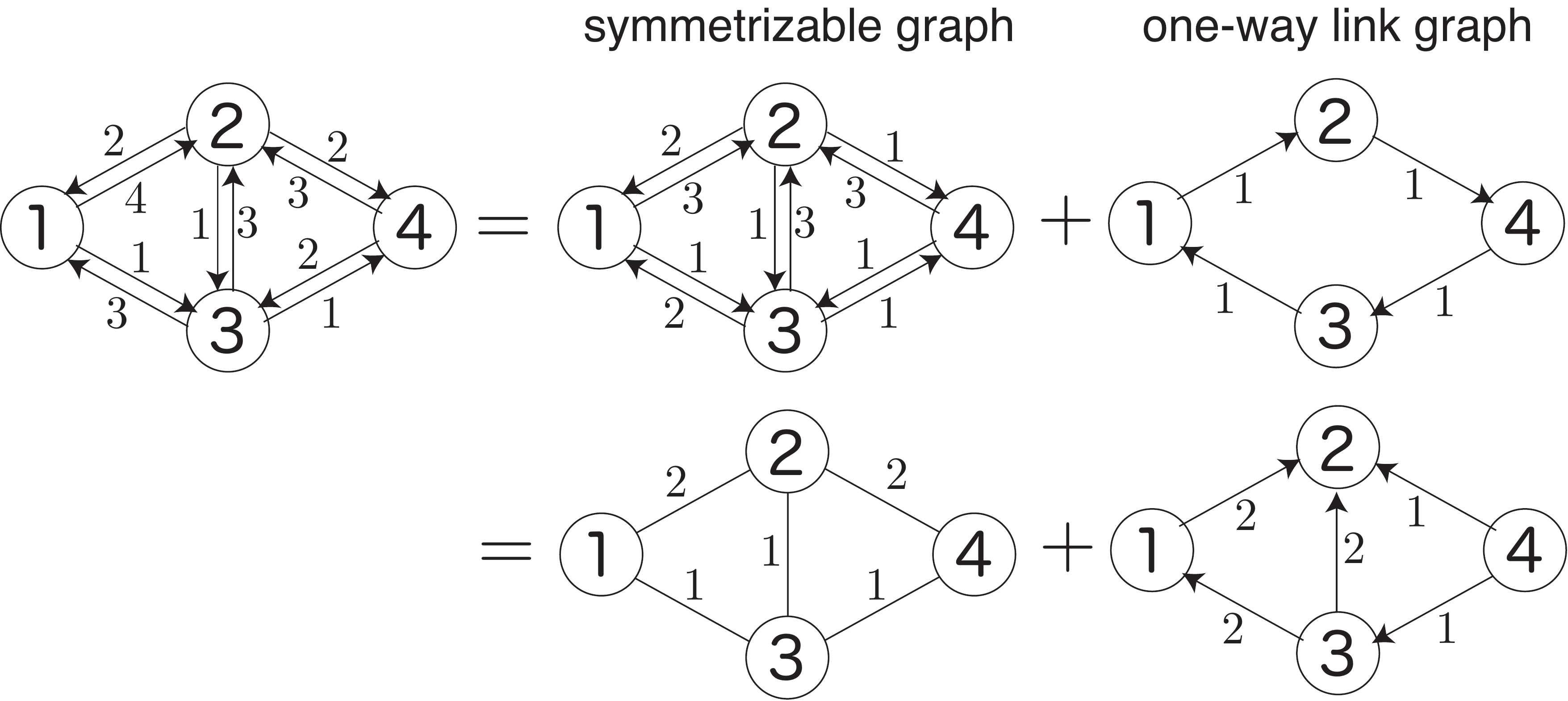}
\caption{Examples of the decomposition $\bm{\mathcal{L}} = \bm{\mathcal{L}}_0 + \bm{\mathcal{L}}_\mathrm{I}$}
\label{fig:decomp-0} 
\end{center}
\end{figure}

Figure~\ref{fig:decomp-0} shows examples of decomposition (\ref{eq:L-decomp}) of a directed graph. 
As shown in the first decomposition of Fig.~\ref{fig:decomp-0}, the directed graph on the left-hand side is composed of the symmetrizable directed graph depicted in Fig.~\ref{fig:ex} and a one-way link graph. 
The second decomposition of Fig.~\ref{fig:decomp-0} yields an undirected graph and a one-way link graph. 
The undirected graph is a special case of symmetrizable directed graphs; $\bm{M}=\bm{I}$ (unit matrix) in this case.

From the above example, we can recognize the following facts:
\begin{itemize}
\item Any directed graph can be decomposed into a symmetrizable directed graph and a one-way link graph. 
\item The decomposition is not unique, as it depends on the choice of symmetrizable directed graph included in the original directed graph. 
\end{itemize}
If the original directed graph is symmetrizable, we can choose a decomposition such that $\bm{\mathcal{L}}_\mathrm{I}=\bm{\mathrm{O}}$ (the null matrix). 

\subsection{Online Flaming and Critical phenomena}
From (\ref{eq:E_i}), the oscillation energy is time-independent for symmetrizable directed graphs, so no flaming can occur in any symmetrizable directed graph. 
Let us consider a general directed graph by adding a one-way link graph to a symmetrizable directed graph.  
Here, let the strength of the influence of one-way link graphs be indicated by a parameter, $\epsilon \ge 0$, and we introduce Laplacian matrix $\bm{\mathcal{L}}(\epsilon)$ for general directed graphs as 
\begin{align}
\bm{\mathcal{L}}(\epsilon) = \bm{\mathcal{L}}_0 + \epsilon \, \bm{\mathcal{L}}_\mathrm{I}. 
\label{eq:L=L_0+L_I}
\end{align}
The value of $\epsilon$ indicates the magnitude of deviation from the symmetrizable directed graph, and $\bm{\mathcal{L}}(0) = \bm{\mathcal{L}}_0$. 

It is worth noting that we do not assume the structure of OSN should be a symmetrizable graph.
Our purpose here is to investigate the difference in user dynamics depending on the OSN structures. 
Here, the structure is parametrized by $\epsilon$ and the value of $\epsilon$ is used as an index representing the deviation from a symmetrizable graph.
In the rest of this subsection, we show a preliminary evaluation of user dynamics using an example network model. 

\begin{figure}[bt]
\begin{center}%
 \includegraphics[width=0.85\linewidth]{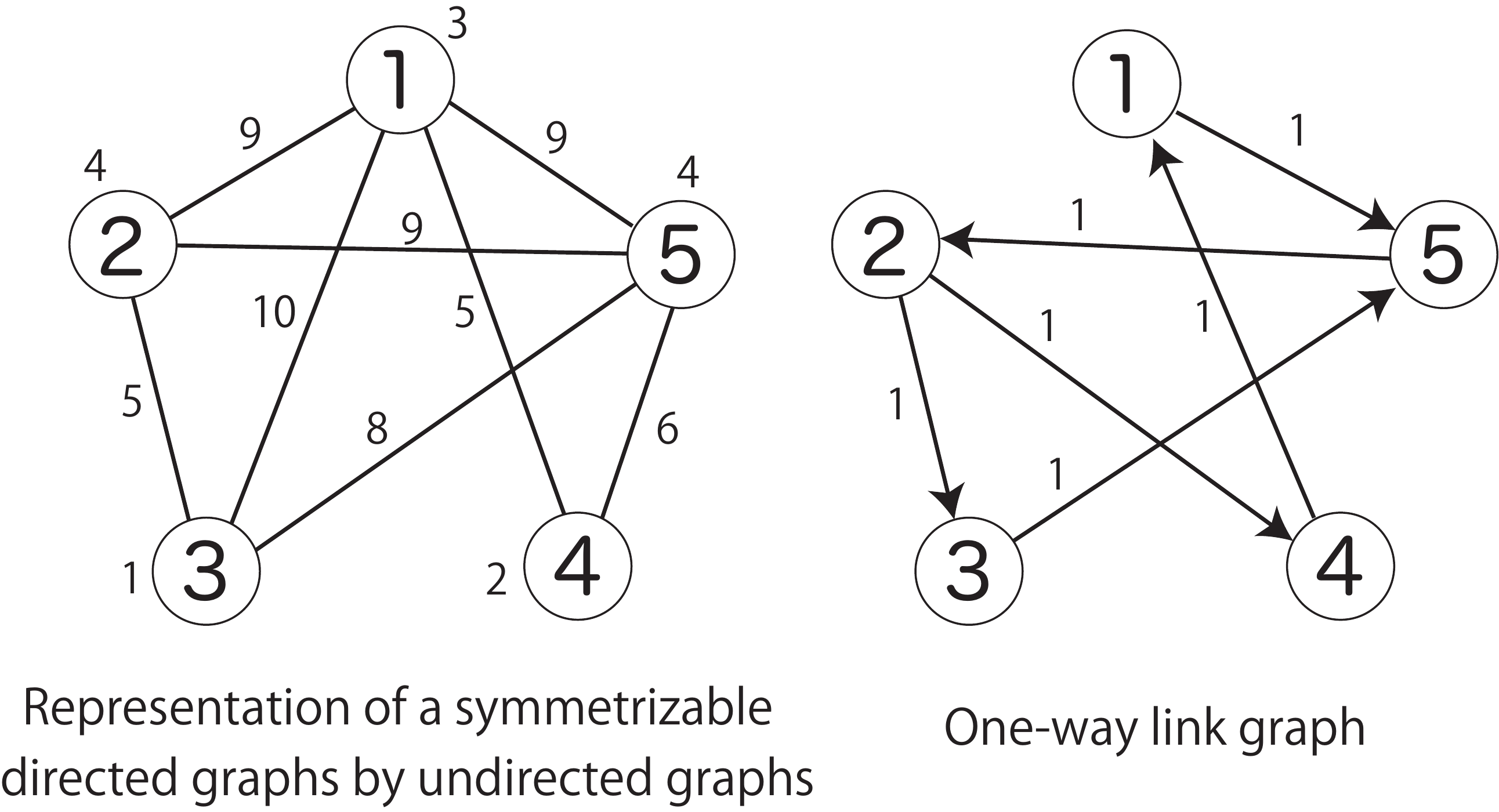}
\caption{Directed network model composed of a symmetrizable directed graph and a one-way link graph}
\label{fig:NW-model-omen}
\end{center}
\end{figure}

As an example, we consider the following network model. 
The symmetrizable directed graph is given as 
\begin{align}
\bm{\mathcal{L}}_0 &= 
\begin{bmatrix}
11 & -3 & -10/3 & -5/3 & -3 \\
-9/4 & 23/4 & -5/4 & 0 & -9/4\\
-10 & -5 & 23 & 0 & -8\\
-5/2 & 0 & 0 & 11/2 & -3\\
-9/4 & -9/4 & -2 & -3/2 & 8
\end{bmatrix}
\notag\\
%
&= 
\begin{bmatrix}
3 & 0 & 0 & 0 & 0 \\
0 & 4 & 0 & 0 & 0\\
0 & 0 & 1 & 0 & 0\\
0 & 0 & 0 & 2 & 0\\
0 & 0 & 0 & 0 & 4
\end{bmatrix}^{-1}
\begin{bmatrix}
33 & -9 & -10 & -5 & -9 \\
-9 & 23 & -5 & 0 & -9\\
-10 & -5 & 23 & 0 & -8\\
-5 & 0 & 0 & 11 & -6\\
-9 & -9 & -8 & -6 & 32
\end{bmatrix}, 
\notag
\end{align}
and the one-way link graph is given as 
\begin{align}
\bm{\mathcal{L}}_\mathrm{I} &= 
\begin{bmatrix}
1 & 0 & 0 & 0 & -1 \\
0 & 2 & -1 & -1 & 0\\
0 & 0 & 1 & 0 & -1\\
-1 & 0 & 0 & 1 & 0\\
0 & -1 & 0 & 0 & 1
\end{bmatrix}. 
\notag
\end{align}
Figure~\ref{fig:NW-model-omen} illustrates them. 
The left panel of Fig.~\ref{fig:NW-model-omen} shows the symmetrizable directed graph depicted in the symmetrization manner similar that used in Fig.~\ref{fig:ex2} and the right panel shows the one-way link graph.  

\begin{figure}[bt]
\begin{center}%
 \includegraphics[width=0.95\linewidth]{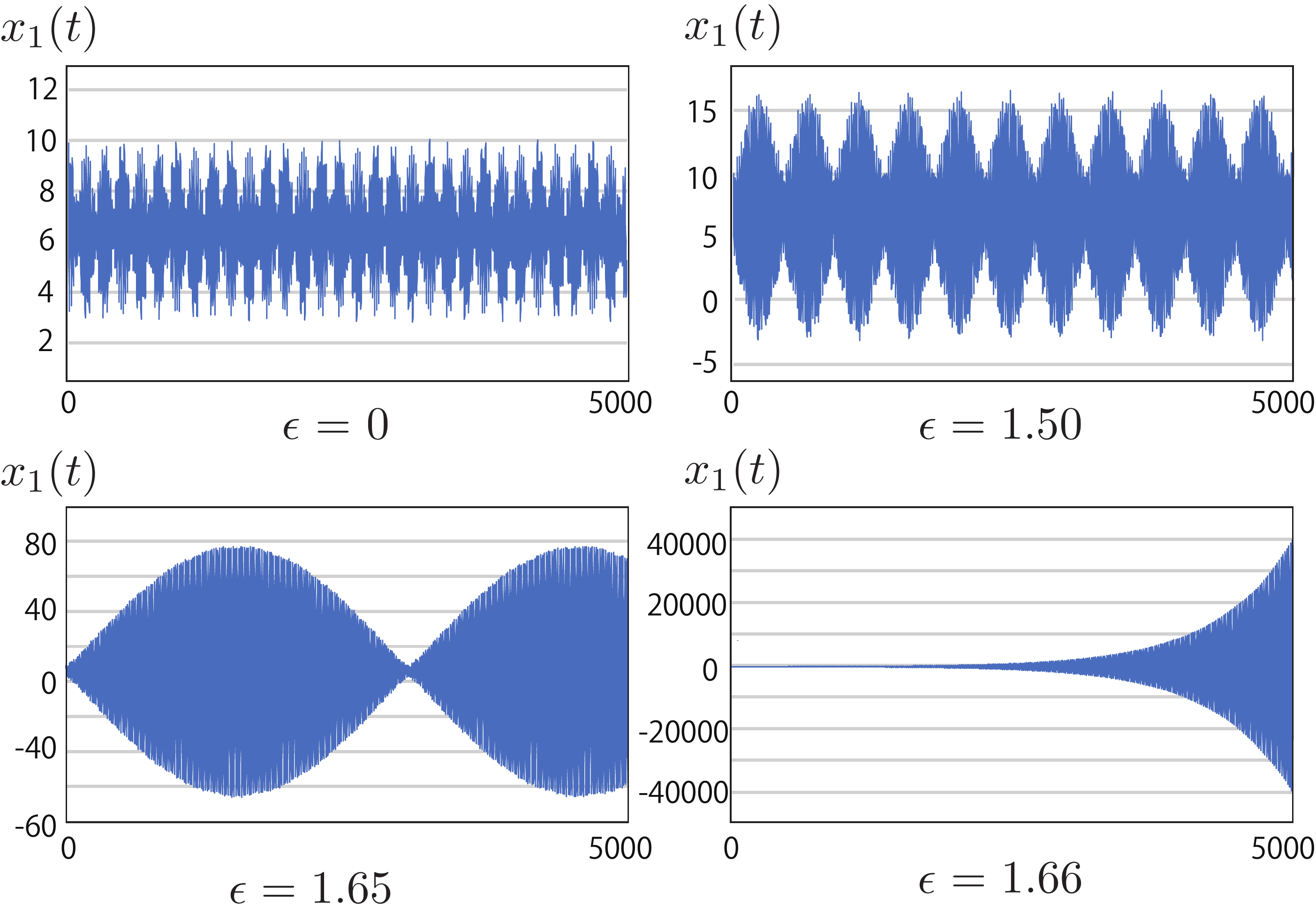}
\caption{Time variation of displacement $x_1(t)$ of node $1$ for directed graphs of various values of $\epsilon$}
\label{fig:beat->diverge}
\end{center}
\end{figure}

We demonstrate a typical example of user dynamics by using this network model. 
Fig.~\ref{fig:beat->diverge} shows the time variation of user state $x_1(t)$ for node $1$ given the $\epsilon =  0$, $1.5$, $1.65$ and $1.66$. 
The initial vector is set to $\bm{x}(0) = {}^t\!(10,\,2,\,7,\,5,\,6)$. 
The boundary value of the parameter at which non-real eigenvalues appear lies in the range of $1.65 < \epsilon < 1.66$. 
For $\epsilon>0$, the network model is not symmetrizable but all the eigenvalues are real for $\epsilon<1.65$. 
For $\epsilon>1.66$, non-real eigenvalues cause divergence. 
From these results, it can be seen that as the value of $\epsilon$ increases, low-frequency beats occur and grow, before non-real eigenvalues appear. 
This phenomenon is not limited to the above example. 
The reason for this is explained in the next subsection.

It is natural to think that OSN structure changes over time.
The situation being discussed here is that changing the structure of the OSN by increasing the value of $\epsilon$ causes the change of the user dynamics to be activated. 
So, even if it does not lead to online flaming, we can expect low-frequency beats to appear when online user dynamics are activated. 

Note that symmetrizable directed networks never diverge, but its converse is not always true. 
For $0 < \epsilon < 1.65$, the OSN structure is not symmetrizable but all the eigenvalues are real and no divergence occurs. 
Moreover, although actual OSNs should exert some damping effect on user dynamics (that is, the effect of calming user dynamics over time)~\cite{aida2018,Takano-access}, the numerical example described here does not consider this effect, for simplicity of explanation. 
If the strength of the damping effect is sufficiently large, user dynamics do not diverge even if some eigenvalues of OSN become non-real. 

\subsection{Why Low-Frequency Mode Appears}
\begin{figure}[bt]
\begin{center}%
 \includegraphics[width=0.95\linewidth]{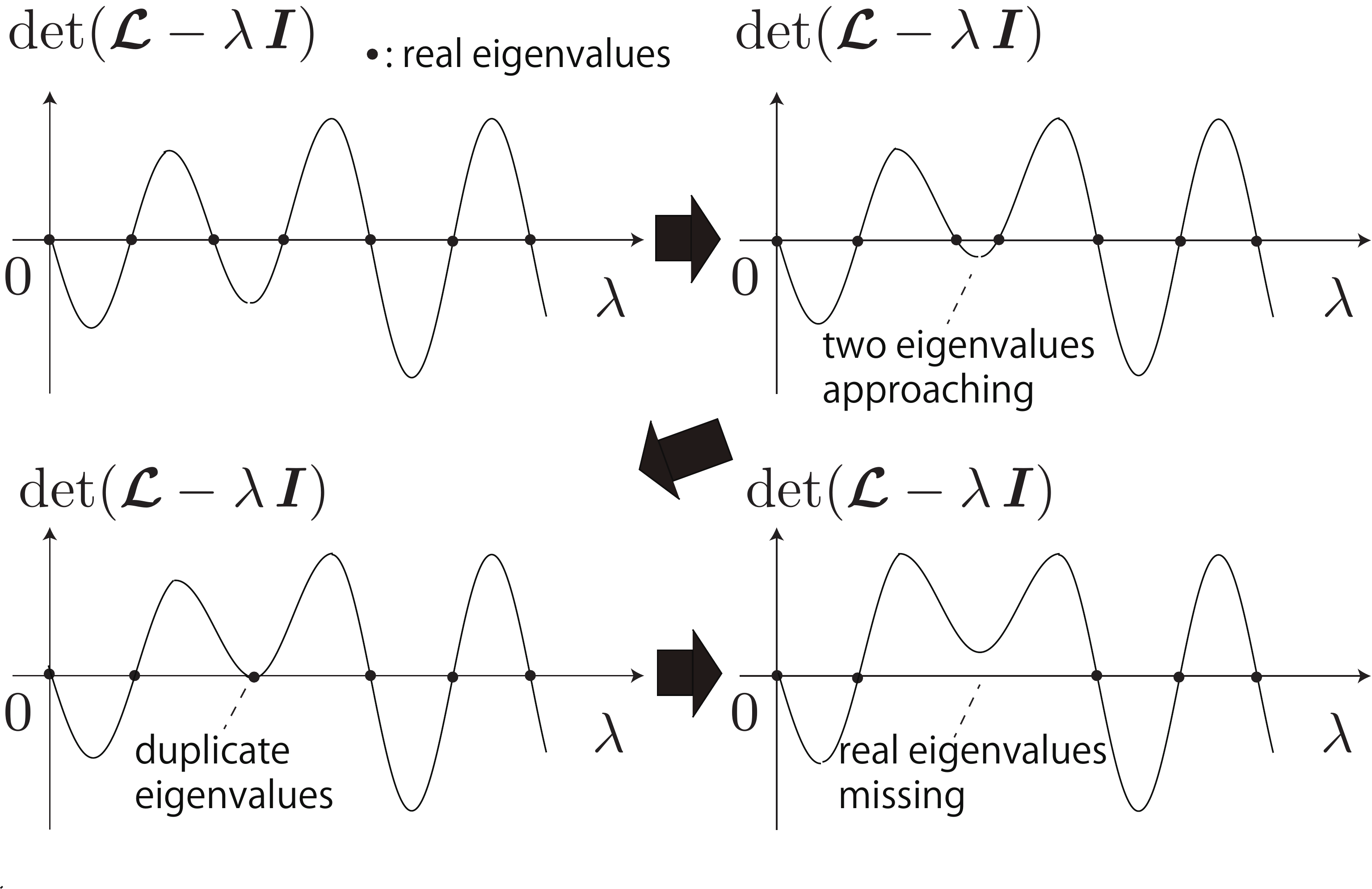}
\caption{Changes in the shape of the characteristic polynomial $\mathrm{det}(\bm{\mathcal{L}}(\epsilon) - \lambda \, \bm{I})$ when $\epsilon$ is increased: Two eigenvalues approach each other before the solutions of the characteristic equation become non-real numbers.}
\label{fig:flaming_eigenvalue}
\end{center}
\end{figure}

The occurrence and growth of the low-frequency beats can be explained by the oscillation model. 
Let us consider the characteristic function of Laplacian matrix $\bm{\mathcal{L}}(\epsilon)$ for $\epsilon$; 
\[
\mathrm{det}(\bm{\mathcal{L}}(\epsilon) - \lambda \, \bm{I}) = 0. 
\]
For $\epsilon = 0$, since $\bm{\mathcal{L}}(0)=\bm{\mathcal{L}}_0$ is a symmetrizable directed graph, all the eigenvalues are real. 
The upper-left panel of Fig.~\ref{fig:flaming_eigenvalue} shows this situation. 
The solutions of the characteristic equation correspond to the intersections with the horizontal axis.
Changes in the shape of the characteristic polynomial $\mathrm{det}(\bm{\mathcal{L}}(\epsilon) - \lambda \, \bm{I})$ when $\epsilon$ is increased are schematically displayed in the order of the arrows.
The last state is where the intersection with the horizontal axis has disappeared and non-real eigenvalues have appeared.
Before reaching that state, two eigenvalues begin to approach each other on the horizontal axis; they disappear immediately after they intersect.
Since an eigenfrequency is the square root of its eigenvalue, the two eigenfrequencies approach when the corresponding eigenvalues approach.
The approach of the two eigenfrequencies triggers the low-frequency beats. 

As a familiar example, let us consider the superposition of two oscillations of the angular frequencies $\omega_1$ and $\omega_2$.  
From the well-known formula of trigonometric functions, the superposition can be written as
\[
\sin(\omega_1\,t) + \sin(\omega_2\,t) 
= 2\,\sin\left(\frac{\omega_1 + \omega_2}{2}\,t\right)  \cos\left(\frac{\omega_1 - \omega_2}{2}\,t\right). 
\]
If the angular frequencies $\omega_1$ and $\omega_2$ are close to each other, $\omega_1-\omega_2$ becomes a very small value. 
A wave oscillating at the average angular frequency $(\omega_1 + \omega_2)/2$ appears as a waveform modulated by a low-frequency wave with angular frequency $(\omega_1-\omega_2)/2$. 

Next, we consider why the amplitude of the low-frequency mode grows. 
Since $\epsilon = 0$ gives a symmetrizable directed graph, the scaled Laplacian matrix $\bm{S}_0$ is a symmetric matrix and its eigenvectors $\{\bm {v}_\mu\}_{\mu=0,\,1,\,\dots,\, n-1}$ can be selected as an orthonormal basis. 
The scaled Laplacian matrix for general directed graphs is defined as 
\[
\bm{S}(\epsilon) := \bm{M}^{+1/2} \, \bm{\mathcal{L}}(\epsilon) \, \bm{M}^{-1/2}. 
\]
For $\epsilon > 0$, $\bm{S}(\epsilon)$ is an asymmetric matrix. 

\begin{figure}[bt]
\begin{center}%
 \includegraphics[width=0.95\linewidth]{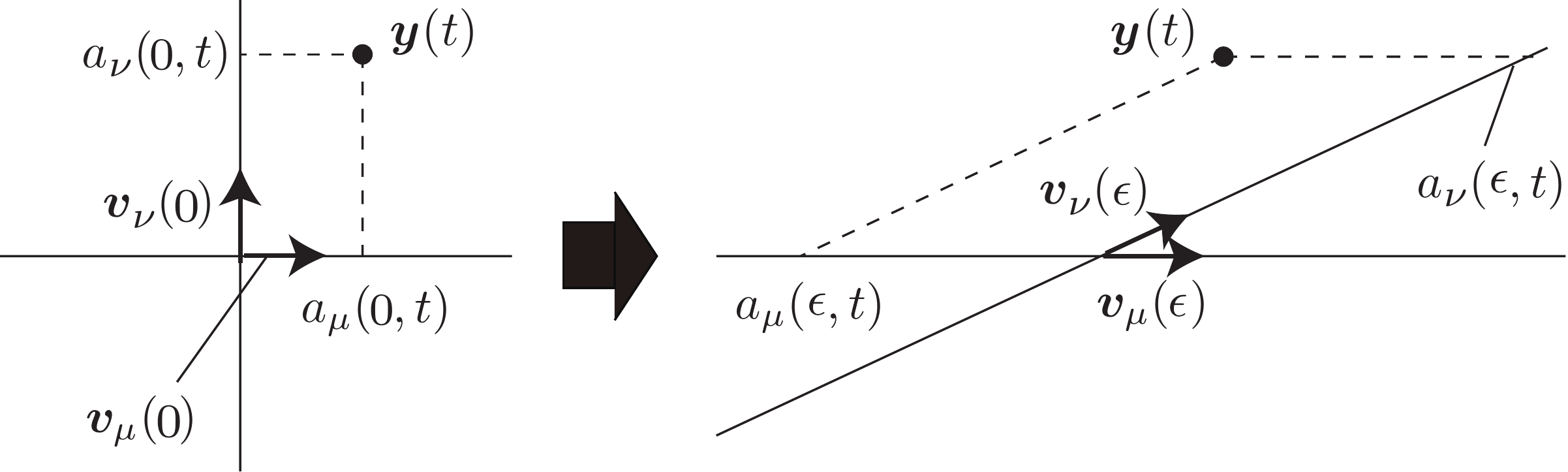}
\caption{Expansion coefficient increases as two eigenvectors become parallel}
\label{fig:flaming_eigenvector}
\end{center}
\end{figure}

Thus the eigenvectors $\{\bm {v}_\mu(\epsilon)\}_{0\le\mu\le n-1}$ of $\bm{S}(\epsilon)$ are not orthogonal to each other. 
For solution $\bm{y}(t)$ of the wave equation 
\[
\frac{\dd^2}{\dd t^2} \, \bm{y}(t) = - \bm{S}(\epsilon) \, \bm{y}(t), 
\]
we expand $\bm{y}(t)$ by the (non-orthogonal but linearly independent) eigenvectors as 
\[
\bm{y}(t) = \sum_{\mu=0}^{n-1} a_\mu(\epsilon,t) \, \bm{v}_\mu(\epsilon). 
\]
Here, the expansion coefficients $a_\mu(\epsilon,t)$ are the coordinate values of the space spanned by the oblique base vectors $\{\bm {v}_\mu(\epsilon)\}_{0\le\mu\le n-1}$.
Figure~\ref{fig:flaming_eigenvector} schematically shows the coordinate value of each axis, which is expanded in an orthogonal coordinate system, and the change in the coordinate value of each axis, which is expanded on an oblique axis. 
As two eigenvalues approach each other as shown in Fig.~\ref{fig:flaming_eigenvalue}, the angle between the two eigenvectors associated with the two eigenvalues becomes smaller.
At the point of the two eigenvalues intersect, the corresponding eigenvectors become parallel. 
Although $\bm{x}(t)$ is a point at the same position relative to the origin, the absolute value of the coordinate value $a_\mu(\epsilon, t)$ increases on the oblique axes.
This is why the amplitude of the low-frequency mode grows when the eigenvalues approach each other. 

The low-frequency beats appear not only in the user state $x_i(t)$ but also in the oscillation energy of the OSN. 
According to \cite{aida-book2020}, if we choose 
\[
\bm{y}(t) = \sum_{\mu=0}^{n-1} |a_\mu(\epsilon,t)| \, \e^{-\omega_\mu t}\, \bm{v}_\mu(\epsilon), 
\]
for simplicity, the oscillation energy of the whole OSN is obtained as 
\begin{align}
E &= \frac{1}{2} \left\|\sum_{\mu=0}^{n-1} \frac{\dd \bm{y}_\mu(t)}{\dd t} \right\|^2
\notag\\
&= \frac{1}{2} \sum_{\mu=0}^{n-1} \sum_{\nu=0}^{n-1} |a_\mu(\epsilon,t)| \, |a_\nu(\epsilon,t)| \, \omega_\mu \, \omega_\nu  \, (\bm{v}_\mu(\epsilon) \cdot \bm{v}_\nu(\epsilon))
\notag\\
&\quad \times 
\exp\!\left[\,-\ii \,(\omega_\mu - \omega_\nu) \, t\right] 
\notag\\
&= \frac{1}{2} \sum_{\mu=0}^{n-1} |a_\mu(\epsilon,t)|^2 \, \omega_\mu^2 
\notag\\
&\quad {}+ \sum_{\mu=0}^{n-1} \sum_{\nu=\mu+1}^{n-1} |a_\mu(\epsilon,t)| \, |a_\nu(\epsilon,t)| \, \omega_\mu \, \omega_\nu  \, (\bm{v}_\mu(\epsilon) \cdot \bm{v}_\nu(\epsilon))
\notag\\
&\qquad \times 
\cos\!\left[\,(\omega_\mu - \omega_\nu) \, t\right], 
\label{eq:energy}
\end{align}
where $|a_\mu(\epsilon,t)|$ denotes the amplitude of $a_\mu(\epsilon,t)$. 
This means that the oscillation energy is time-dependent if the eigenvectors $\{\bm {v}_\mu(\epsilon)\}_{0\le\mu\le n-1}$ are not orthogonal to each other, that is $\bm{v}_\mu(\epsilon) \cdot \bm{v}_\nu(\epsilon) \not= 0$ for some $\mu$ and $\nu$. 
Therefore, the low-frequency beats appear also in the oscillation energy, if the eigenfrequencies approach each other. 

\begin{figure}[bt]
\begin{center}
\includegraphics[width=0.95\linewidth]{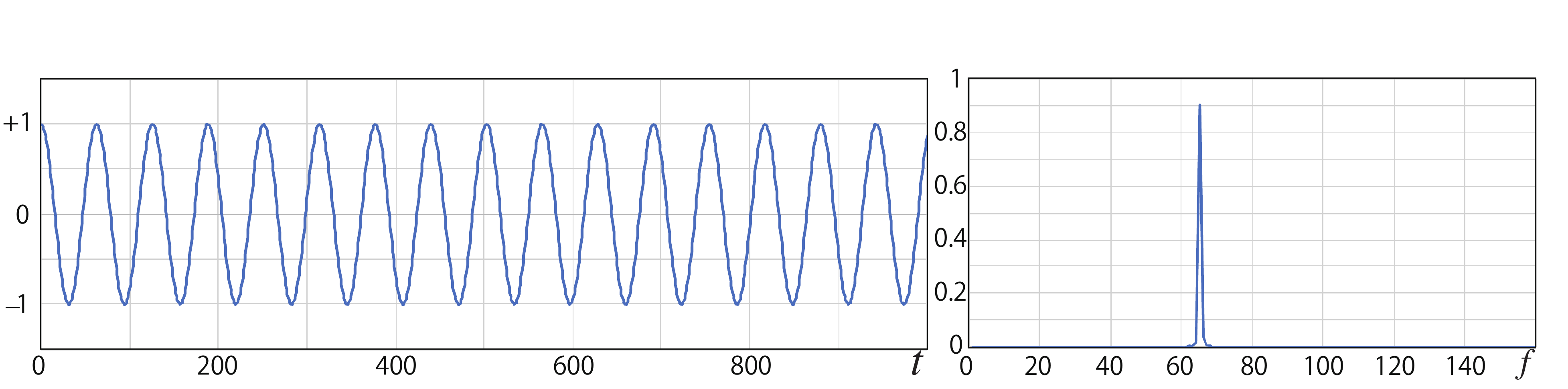}\\
(a) $\cos(0.10 \, t)$ and its spectral distribution\\ 
\includegraphics[width=0.95\linewidth]{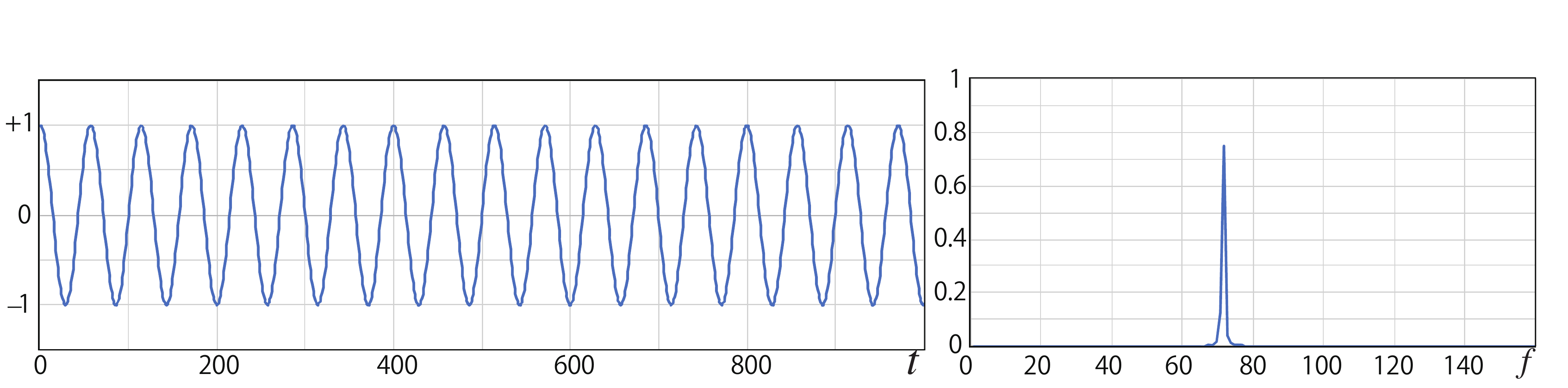}\\
(b) $\cos(0.11 \, t)$ and its spectral distribution\\ 
\includegraphics[width=0.95\linewidth]{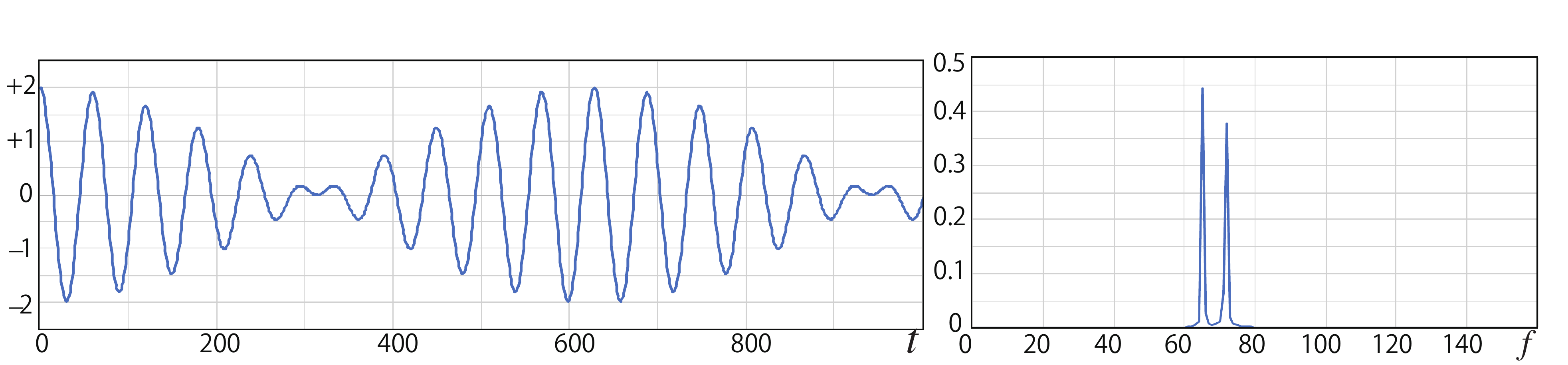}\\
(c) $\cos(0.10 \, t)+\cos(0.11 \, t)$ and its spectral distribution\\ 
\includegraphics[width=0.95\linewidth]{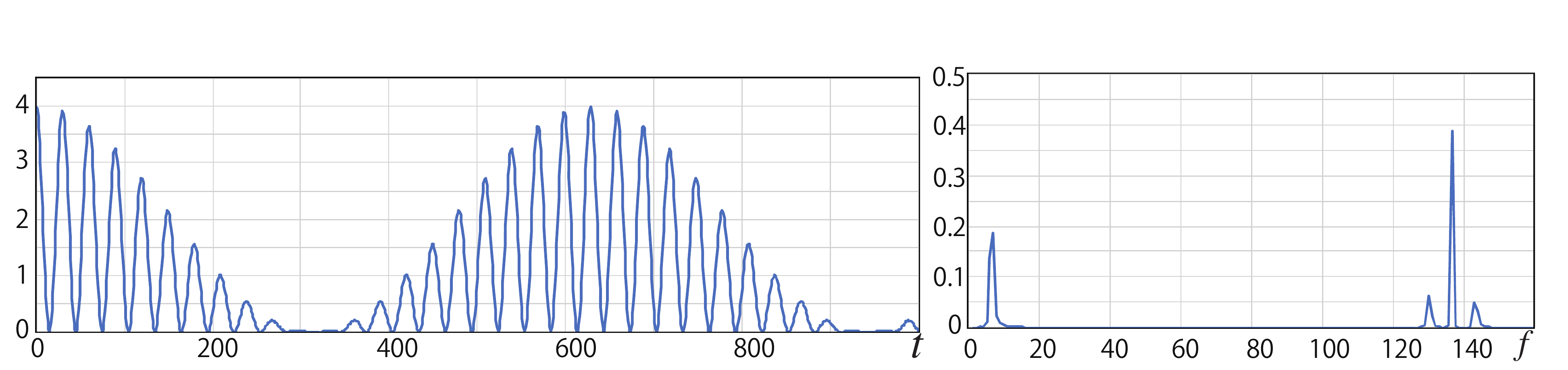}\\
(d) $(\cos(0.10 \, t)+\cos(0.11 \, t))^2$ and its spectral distribution\\ 
\includegraphics[width=0.95\linewidth]{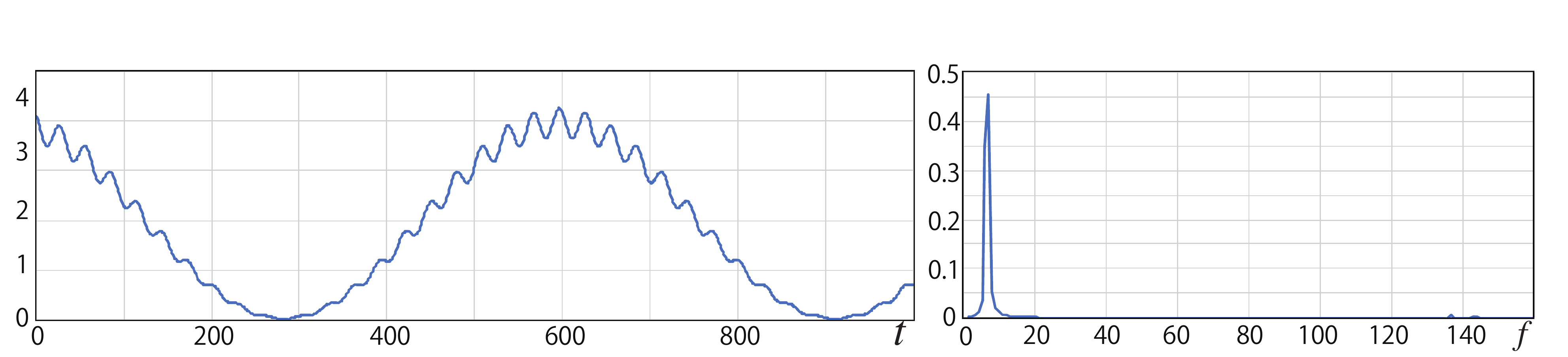}\\
(e) Moving average of $(\cos(0.10 \, t)+\cos(0.11 \, t))^2$ and its spectral distribution (window size: 64)\\ 
\caption{Low-frequency mode appears in the observed oscillation energy}
\label{app:lfb}
\end{center}
\end{figure}

Here, we consider the spectrum in the low-frequency beat of the oscillation energy. 
For simplicity, let us consider two triangular waves $\cos(\omega t)$ with close angular frequencies: $\omega = 0.10$ and $\omega = 0.11$. 
The left panels in Fig.~\ref{app:lfb} (a) and (b) denotes them, respectively. 
Also, their spectral distributions are shown in the right panels in Fig.~\ref{app:lfb} (a) and (b), where the spectral distribution is normalized that the total volume is $1$. Here, the horizontal axes in the right panels denote the frequency $f$ and it is defined as $f := \omega/(2\pi)\times 4096$. 
The peaks of the spectral distribution appear at $f=65.2$ and $f=71.7$, respectively. 
The sum of the two triangular waves is shown in Fig.~\ref{app:lfb} (c). 
A low-frequency beat appears in the left panel, but the spectral distribution has two peaks of individual waves and no low-frequency mode. 

In the oscillation model, the strength of user activity is obtained as the oscillation energy. 
The oscillation energy is related to the square of waves. 
Figure~\ref{app:lfb} (d) shows the square of the sum of the two triangular waves and their spectral distribution. 
The spectral distribution includes mainly two oscillation modes: higher frequency around  $f=137.0 = 65.2 + 71.7$ and lower frequency around $f=6.5 = |65.2 - 71.7|$. 
Here, the lower frequency is the low-frequency beat. 
Moreover, Fig.~\ref{app:lfb} (e) shows a moving average of the square of the sum of the two triangular waves and its spectral distribution. 
The size of the window for the moving average is $64$. 
The meaning to take the moving average is observing the number of specific events during a certain time period. 
The details are shown in the next section. 
Such observation obtains the averaged value of the target in the observing time interval. 
We can recognize that the low-frequency mode is clearly observed in the oscillation energy from the spectral distribution.

Finally, we summarize the framework of observations of low-frequency modes in user activity in OSNs (see Fig.~\ref{app:lfb}). 
The oscillation model is based on the idea that the influence between users propagates at finite speeds and gives the strength of the user's activity as oscillation energy as an observable quantity. 
The wave equation in OSNs is necessary to describe the propagation with a finite speed. 
If we succeed to observe low-frequency beats in the actual user dynamics, it can be said to support the effectiveness of the oscillation model.

\begin{figure}[bt]
\begin{center}
\includegraphics[width=0.8\linewidth]{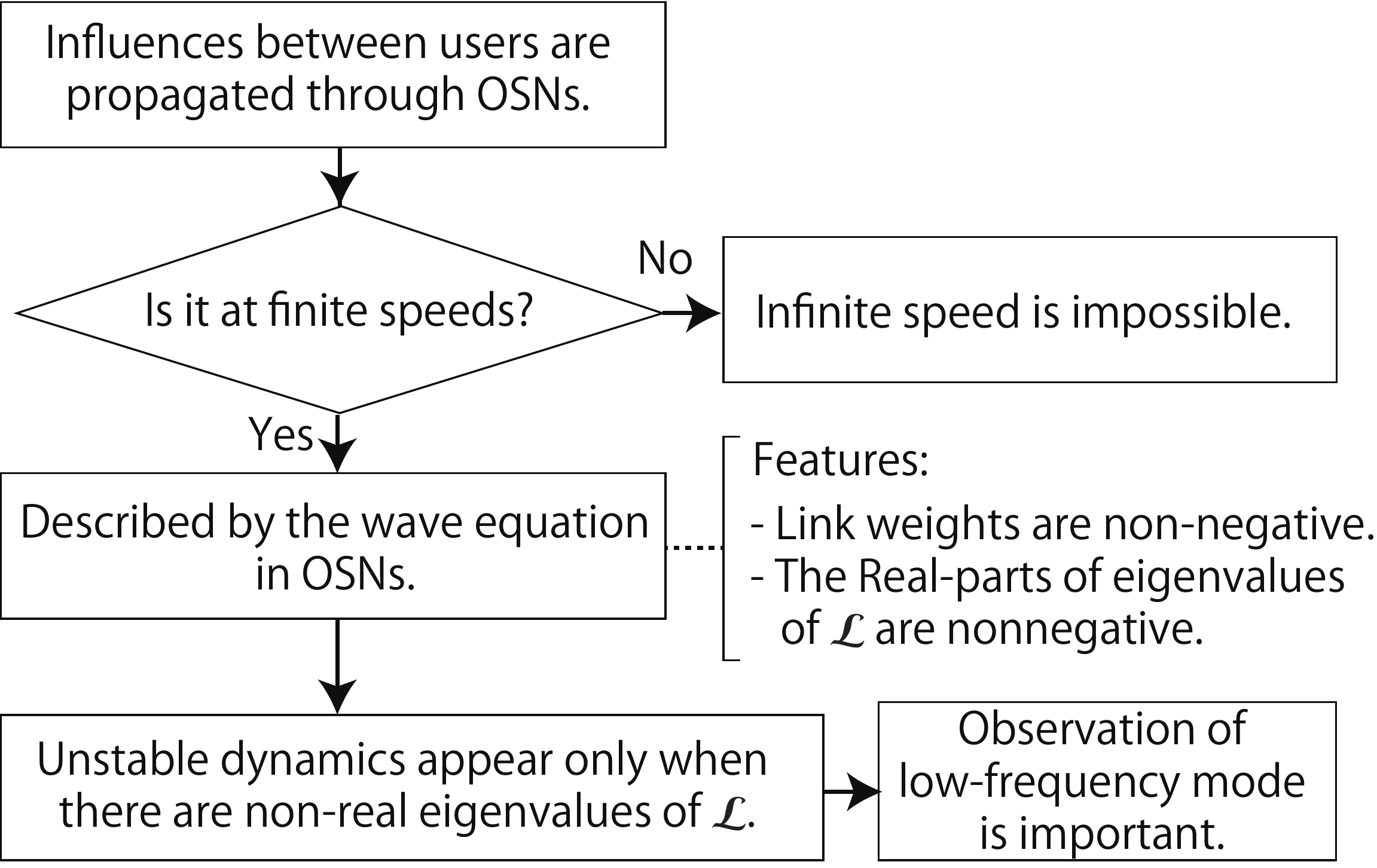}
\caption{Framework of the observation of low-frequency beat}
\label{app:flow_chart}
\end{center}
\end{figure}

\section{Experimental Confirmation of Low-Frequency Mode Occurrence}
\label{sec:exprmnt}
As shown in the previous section, the oscillation model of online user dynamics predicts low-frequency oscillation mode appearance in the strength of user dynamics when the network structure changes to activate user dynamics. 
Accordingly, confirming the appearance of low-frequency modes in real OSN data is considered a reasonable test of the oscillation model's validity. 
This section shows spectral analyses of real data from a bulletin board system (BBS) and Google Trends to confirm the appearance of low-frequency modes.  

In the following analyses, online flaming does not occur, but the intensity of user dynamics grows as the discussion within the OSN change to heat up. 
We verify the occurrence of the low-frequency mode in the highly active periods.

\subsection{Spectral Analysis of BBS data}
2-Channel (the current name is 5-Channel) is the largest BBS in Japan. 
This subsection introduces additional research over our previous study \cite{nagatani}.
In the previous study, we analyzed time-series data sequences of posts to the BBS for different threads. 
Concretely, we performed the spectral analysis by the fast Fourier transformation (FFT) in the period that the number of postings was large (heavy posting period) and the period that the number of postings was small (light posting period) and compare them.
As a result, heavy posting periods exhibited stronger low-frequency modes than light posting periods. 
This result supports the theoretical predictions suggested by the oscillation model. 
However, because \cite{nagatani} does not explain the reasoning behind the selection of threads used to identify heavy/light posting periods, the results presented were of interest but were not definitive. This subsection introduces further experimental results that rectify the shortcomings of \cite{nagatani}. 

We analyzed time-series data sequences of posts to the BBS for two different threads. 
The first thread is a Japanese stock market-related thread, the second one is for the Japanese professional baseball team ``Hiroshima Toyo Carp''. 
The reasons for selecting these threads are as follows. 
Our purpose is to compare the strength of user dynamics in periods in which the discussion becomes animated with that in the periods of a normal state. 
The stock market-related thread is strongly affected by stock price fluctuations. 
When the stock price changes significantly, the discussion in the thread becomes animated, but when the stock price is basically unchanged, the discussion in the thread is not lively. 
Therefore, it is easy to compare the panic-like heavy posting periods with the normal light periods.
Professional baseball is very popular in Japan, and the team ``Hiroshima Toyo Carp'' has been in the spotlight because it won the league title for the third consecutive year. 
The analysis period is the period during and after the Japan series (Japan version of the World Series), including the period in which many fans enthusiastically participated in the discussions.
This choice is also suitable for comparing periods when user debates are animated with those of quiescent discussion.

In addition to the above threads, \cite{nagatani} includes a Korean economics-related thread.  
This is also suitable for selecting heavy and light posting periods. 
However, the selection of multiple heavy and light posting periods different from  \cite{nagatani} is difficult, so we omit it in the following discussion. 

\subsubsection{Thread for Japanese stock market}
\begin{figure}[bt]
\begin{center}
\includegraphics[width=0.85\linewidth]{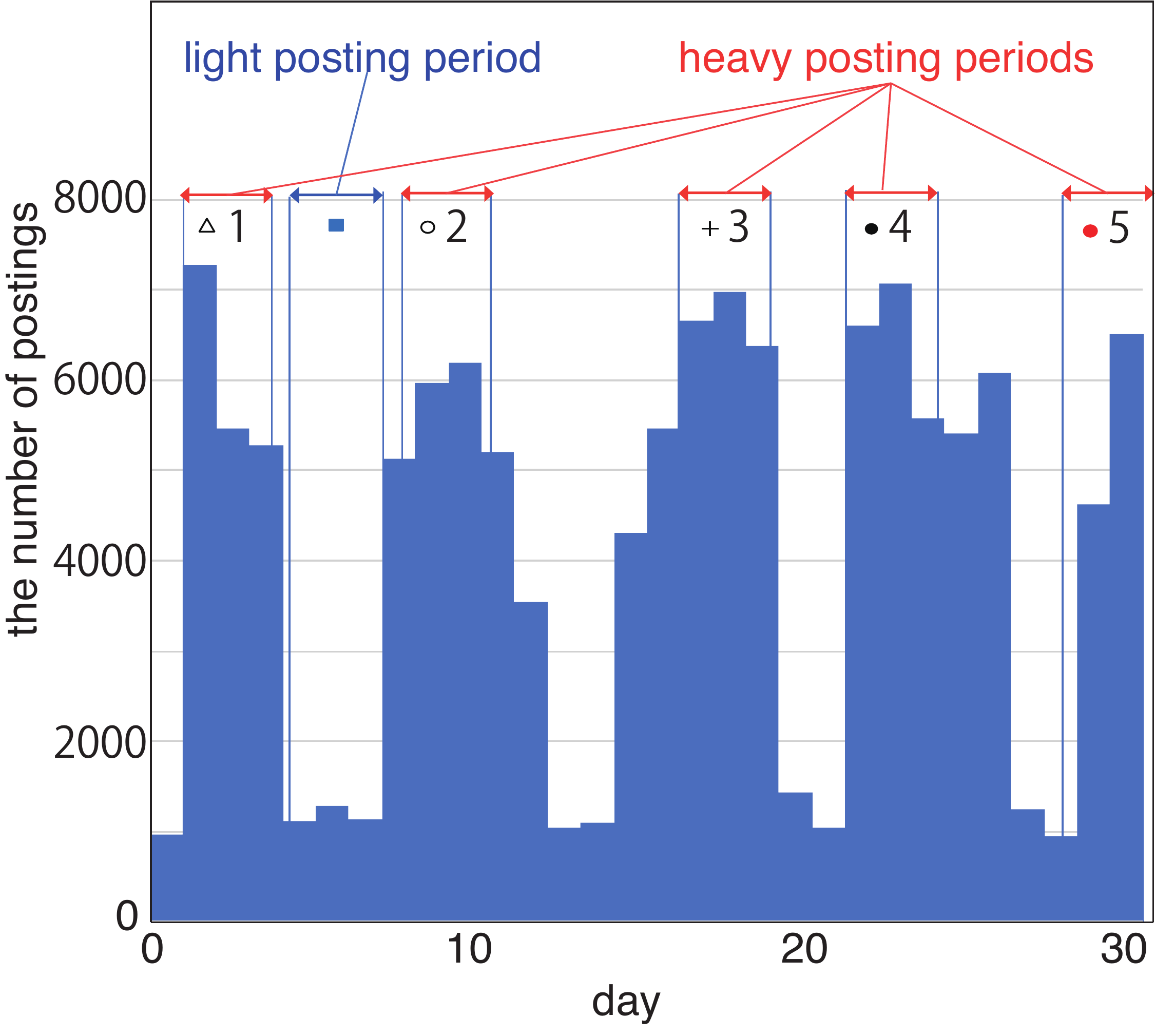}
\caption{The number of posts per day on a Japanese stock market-related thread in 2-Channel}
\label{Japan_stock_256_periods}
\end{center}
\end{figure}

Figure~\ref{Japan_stock_256_periods} shows the number of posts per day from November 19, 2018, to December 7, 2018, made on the Japanese stock market-related thread.
In Fig.~\ref{Japan_stock_256_periods}, the period indicated by a blue arrow has a relatively light-postings period, while that indicated by red arrows have heavy-postings periods. 
The length of each period has successive 256 data long for the number of posting per 16~min. (about 2.84 days long). 
In \cite{nagatani}, only one pair of light and heavy posting periods were analyzed. 
Thus it was possible the period chosen inadvertently matched the hypothesis. 
In this experiment, we chose as many heavy posting periods as we could choose without duplication.
Unfortunately, there is only one light posting period due to the length of its duration; this is discussed later.

\begin{figure}[bt]
\begin{center}
\includegraphics[width=0.85\linewidth]{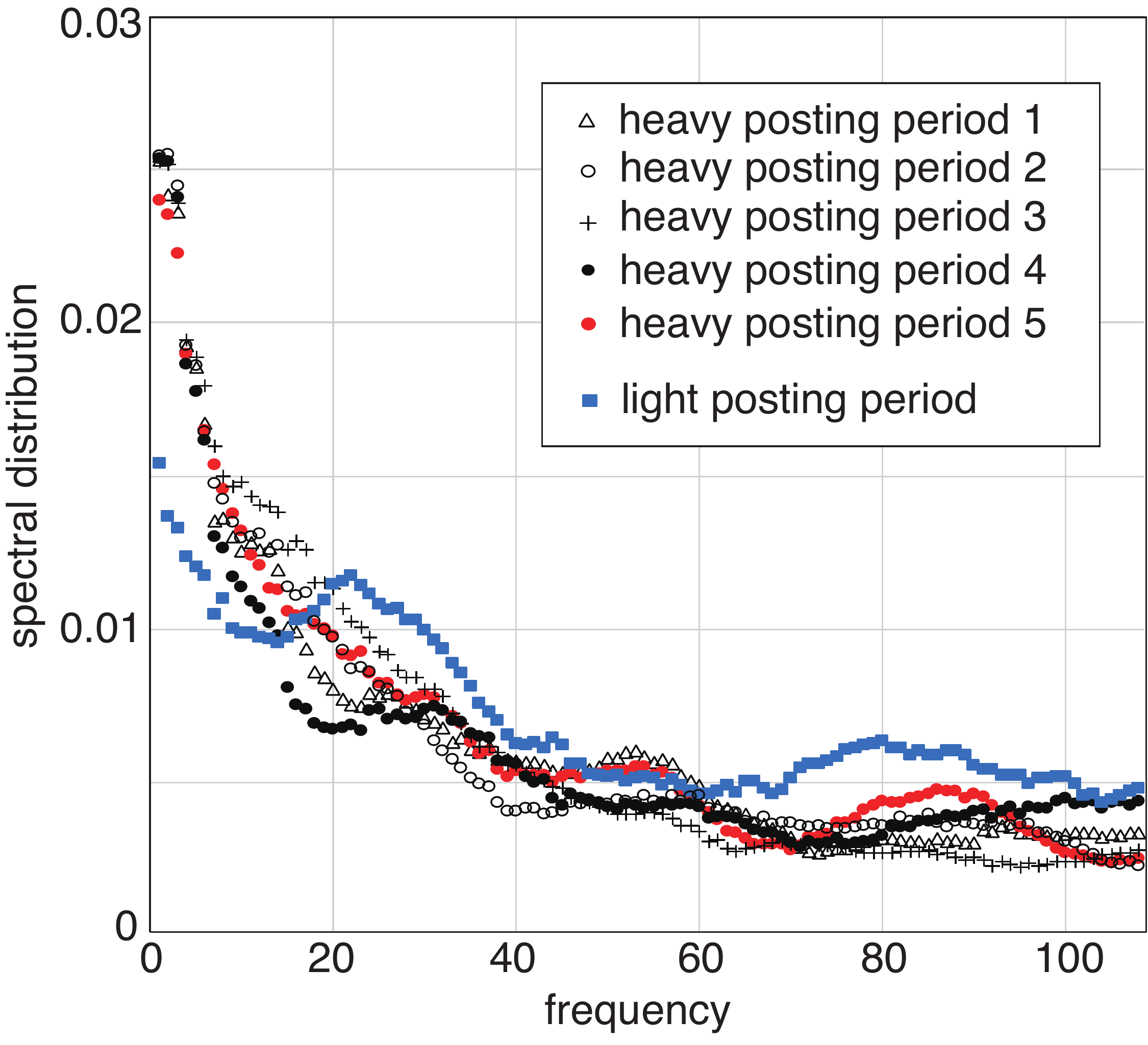}
\caption{Moving average of oscillation mode for window size of $20$ for Japanese stock market-related thread (256 data long)}
\label{Japan_stock_256}
\end{center}
\end{figure}

To compare the spectral distributions for the periods, we processed the data as follows. 
We removed the oscillation mode with a frequency of $0$ to extract only oscillating modes.
Also, to compare the distribution of the ratio of each frequency component, the data were normalized such that the sum of distribution values was $1$.
Fig.~\ref{Japan_stock_256} shows the normalized FFT (moving average with a window size of $20$) of the data in each period. 
Markers indicated in Fig.~\ref{Japan_stock_256} correspond to those in Fig.~\ref{Japan_stock_256_periods}. 
This result shows that the heavy posting periods have stronger low-frequency modes than the light posting period.

It is worth noting that whether the average number of posts in each period is high or low does not impact the spectral analysis. 
They are related to oscillation mode with a frequency of $0$, but they are excluded from the data, and the other oscillation modes are also normalized to be $1$ overall.
This also applies to the subsequent spectral analysis in this paper. 
Moreover, the moving average was executed after the FFT, and the window size of the moving average was selected in order to display the difference in data in an easy-to-see manner.

\begin{figure}[bt]
\begin{center}
\includegraphics[width=0.85\linewidth]{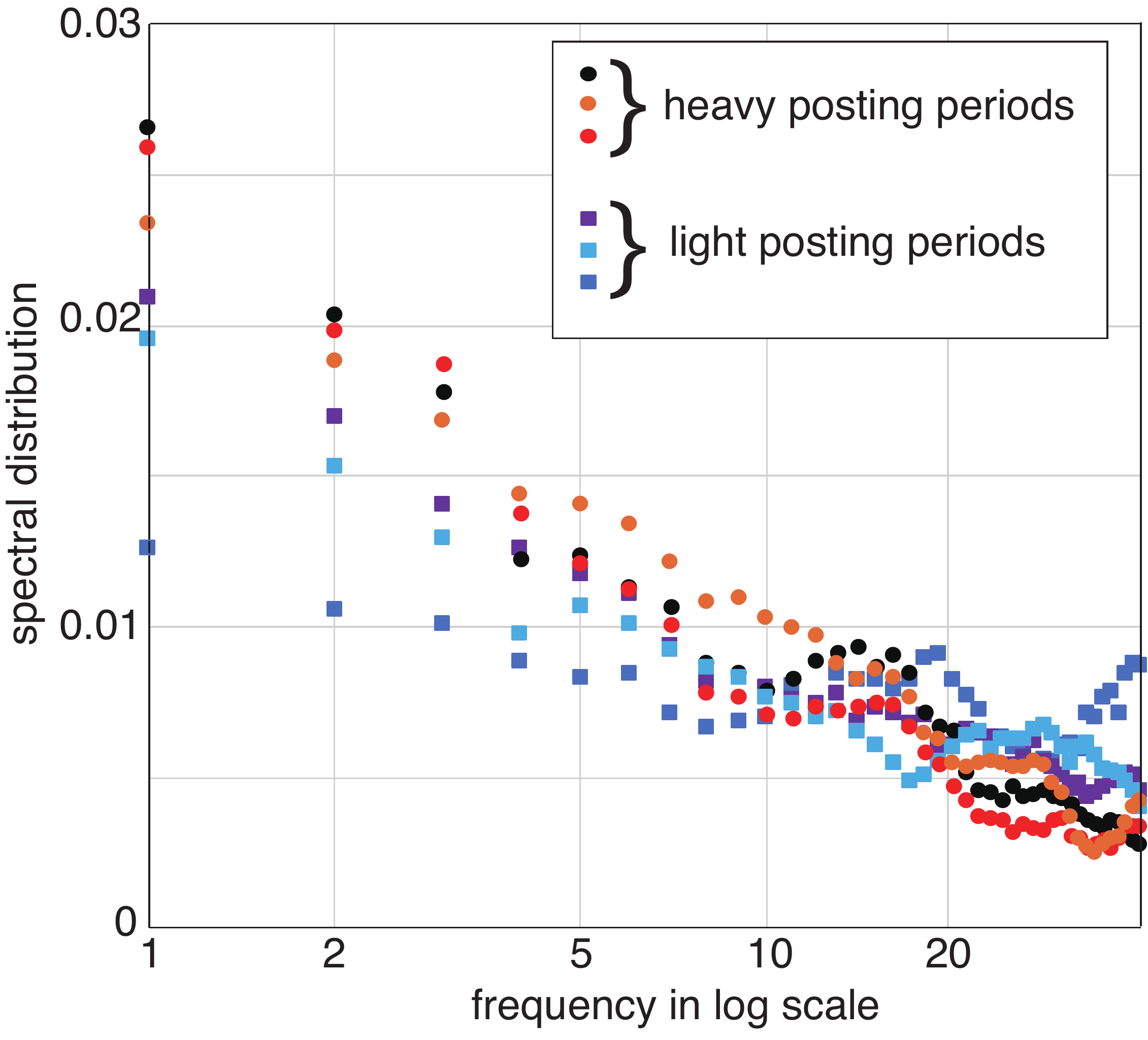}
\caption{Moving average of oscillation mode for window size of $20$ for Japanese stock market-related thread (128 data long)}
\label{Japan_stock_128}
\end{center}
\end{figure}

In the above evaluation, only one sequence of a light posting period could be selected.
To select and evaluate multiple light posting periods, we halved the period length to create 128 contiguous 16 minute periods (total is about 1.42 days).
Under this condition, we select the top 3 non-overlapped periods from the heavy posting periods and the top 3 non-overlapped periods from the light posting period and analyze them. 
Note that this shortening of the data period is disadvantageous in terms of analyzing the low-frequency region. 
Figure~\ref{Japan_stock_128} shows the normalized FFT (moving average with a window size of $20$) of the data in each period. 
To emphasize the low-frequency region, the horizontal axis is plotted using a log scale. 
This result also shows that the heavy posting periods have stronger low-frequency modes than the light posting period.

\subsubsection{Thread for Japanese professional baseball team ``Hiroshima Toyo Carp''} 
\begin{figure}[bt]
\begin{center}
\includegraphics[width=0.85\linewidth]{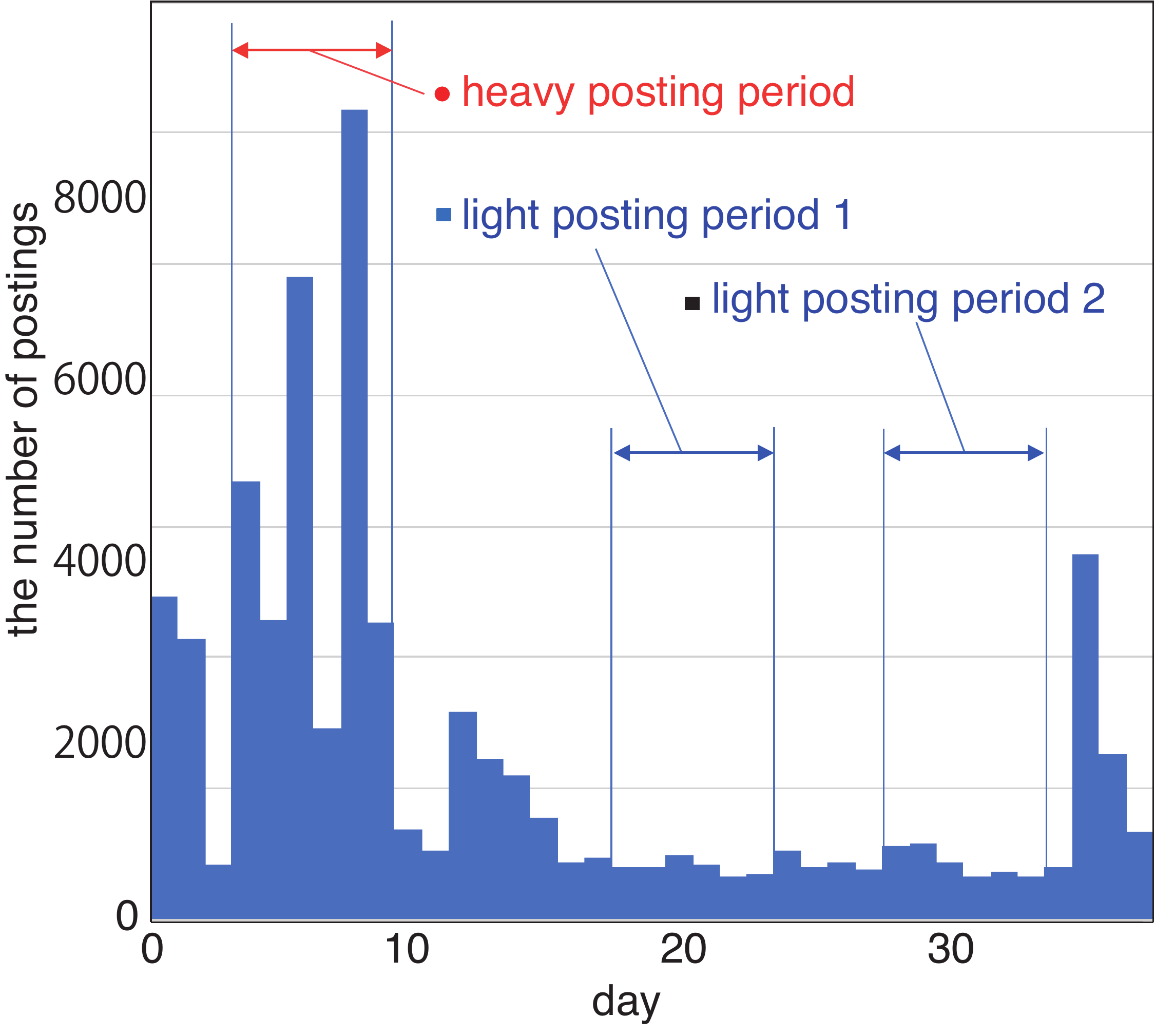}
\caption{The number of posts per day on a Japanese professional baseball team (Hiroshima Toyo Carp) related thread in 2-Channel}
\label{CARP_256_periods}
\end{center}
\end{figure}

Next, Fig.~\ref{CARP_256_periods} shows the number of posts per day from October 27, 2018, to December 3, 2018, made on the Japanese professional baseball team ``Hiroshima Toyo Carp'' related thread.
In Fig.~\ref{CARP_256_periods}, the period indicated by blue arrows have relatively light-postings periods (indicated by blue and black markers), while that indicated by red arrows has a heavy-postings period (indicated by the red marker). 
The length of each period has successive 256 data long for the number of posting per 16~min. (about 2.84 days long). 
In \cite{nagatani}, only one pair of light and heavy posting periods were analyzed. 
In this experiment, we chose two unique light posting periods.

Figure~\ref{CARP_256_periods} shows the normalized FFT (moving average with a window size of $20$) of the data in each period. 
This result also shows that the heavy posting period has stronger low-frequency modes than the light posting period.

\begin{figure}[bt]
\begin{center}
\includegraphics[width=0.85\linewidth]{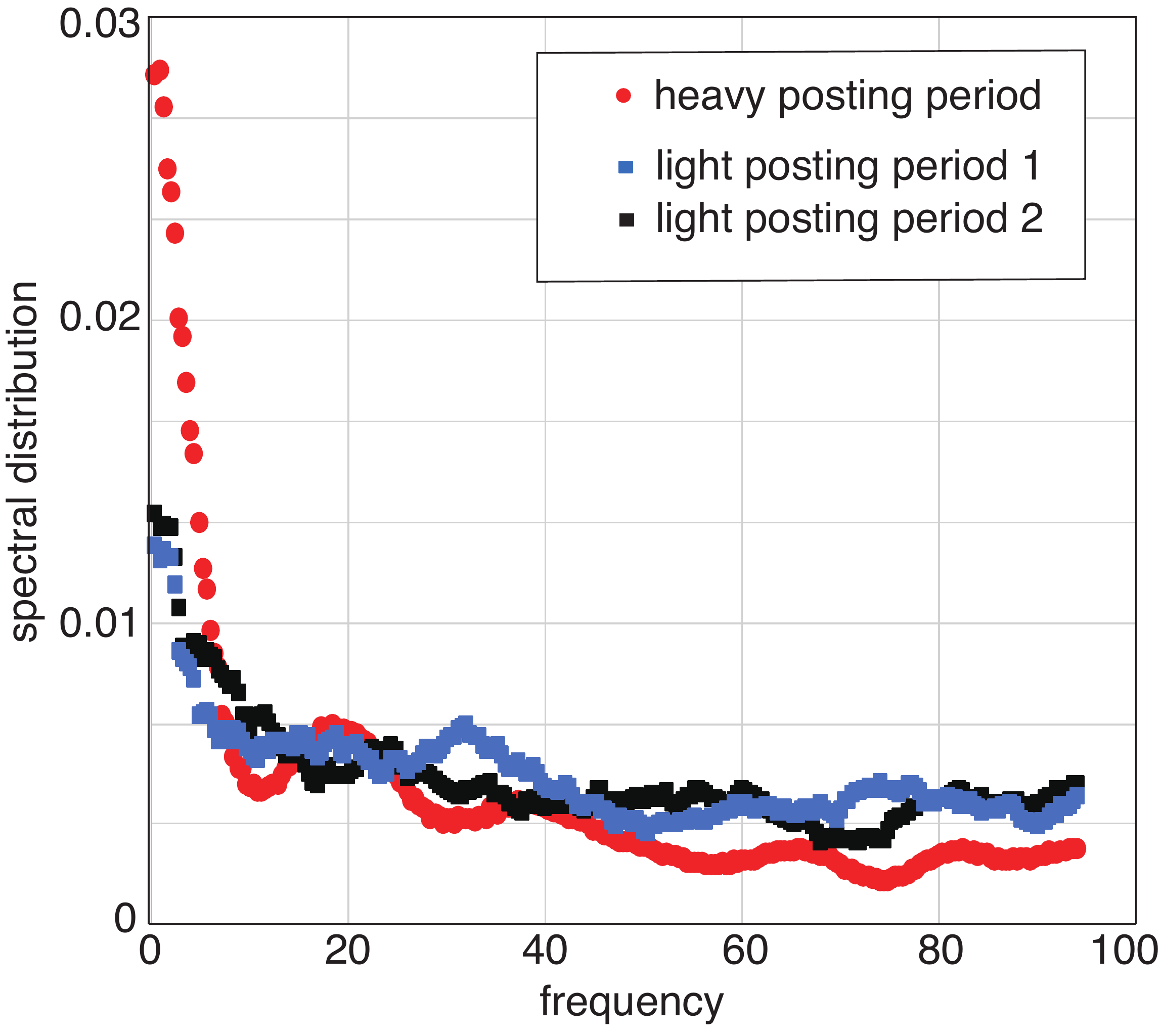}
\caption{Moving average of oscillation mode for window size of $20$ for a Japanese professional baseball team (Hiroshima Toyo Carp) related thread (256 data long)}
\label{CARP_256}
\end{center}
\end{figure}

\subsection{Spectral Analysis of Google Trends Data}
We spectrally analyze time-series data of user interests in words obtained from Google Trends. 
Google Trends is a web service offered by Google that analyzes the popularity of words in Google searches across a certain region in a period. 
For various words such as news keywords attracting attention, we extract the strong interest period and weak interest period and analyze spectral distribution in each period.
The popularity of Google search words does not directly correspond to actual discussion among users in OSNs, but it shows how much OSN interest there is in the topic containing the word. 
Therefore, it is considered that if words associated with particular topics are used in more searches, those topics are attracting significantly more OSN attention, and that word search frequency can be used as an index of the strength of user interest in the related topic.

The data of user interest in a particular word, obtained from Google Trends, is called the search interest.
The specifications of the search interest are summarized as follows.
The search interest is an index that shows how much a word is searched in Google in a certain period.
The data retrievable from Google Trends is limited to a continuous week.
In addition, the values of the search interest are normalized so that the maximum value in the time-series data is $100$.

For this paper, we retrieved multiple hourly time-series data sequences of the search interest (one-week periods) and created long-term data by fusing those one-week data sets.
Consider the example of fusing data set from January 1 to January 7, 2019, and data set from January 7 to January 13, 2019.
Here, if the value of the former at 0:00 January 7 is 80 and the value of the latter at 0:00 on January 7 is 40, the values of the former are multiplied by 0.5.
This approach ensures that the largest value in the fused data is $100$.
The two data sets are then seamlessly combined on the shared date.
In this paper, our target area is Japan.

This subsection details further research that extends our previous study \cite{nagatani_bigdata_2019}.
In the previous study, we analyzed the time-series of the search interest data for some different words of interest. 
As a result, strong interest periods exhibited stronger low-frequency modes than weak interest periods. 
This result supports the theoretical predictions suggested by the oscillation model. 
However, because \cite{nagatani_bigdata_2019} does not discuss the reasons for the selection of the words, further research was deemed necessary. 
We are interested in the periods in which a certain keyword suddenly attracted attention and was discussed. 
Not the characteristic of daily, weekly, or monthly changes in search interest. 
That is, we need some control groups whose search interests change periodically, and to verify that the low-frequency mode does not change in these situations. 
This subsection introduces further experimental results that rectify the shortcomings of \cite{nagatani_bigdata_2019}.

For comparison, we select two words discussed in the previous study \cite{nagatani_bigdata_2019}. 
One is ``7pay'': a Japanese smartphone payment service and the other is ``Fukuoka Softbank Hawks'': a Japanese professional baseball team. 
7pay was an electronic payment system that attracted attention, but due to security problems, there were frequent unauthorized accesses just after the service started, and the service was eventually abolished.
Since this incident became big news and attracted much attention, the periods before and after the incident were analyzed. 
The Japanese professional baseball team, Fukuoka Softbank Hawks, defeated the aforementioned Hiroshima Toyo Carp in the Japanese series and became the Japanese champion.
Since the next season of Fukuoka Softbank Hawks was getting a lot of attention, the periods before and after the start of the next season were analyzed. 
In both words (and the remaining words selected in \cite{nagatani_bigdata_2019}), fluctuations in search interest are not routine, but strongly dependent on topical news.

In addition to the above two words, we select the following three words; ``fgo'': a smartphone game, ``shuukatsu'': job hunting, and ``denshi kessai'': electronic payment.  
They exhibited periodic changes in search interest and their trends have decreasing, increasing, and constant tendencies, respectively. 

\subsubsection{``7pay'': a Japanese smartphone payment service}
\begin{figure}[bt]
\begin{center}
\includegraphics[width=0.48\linewidth]{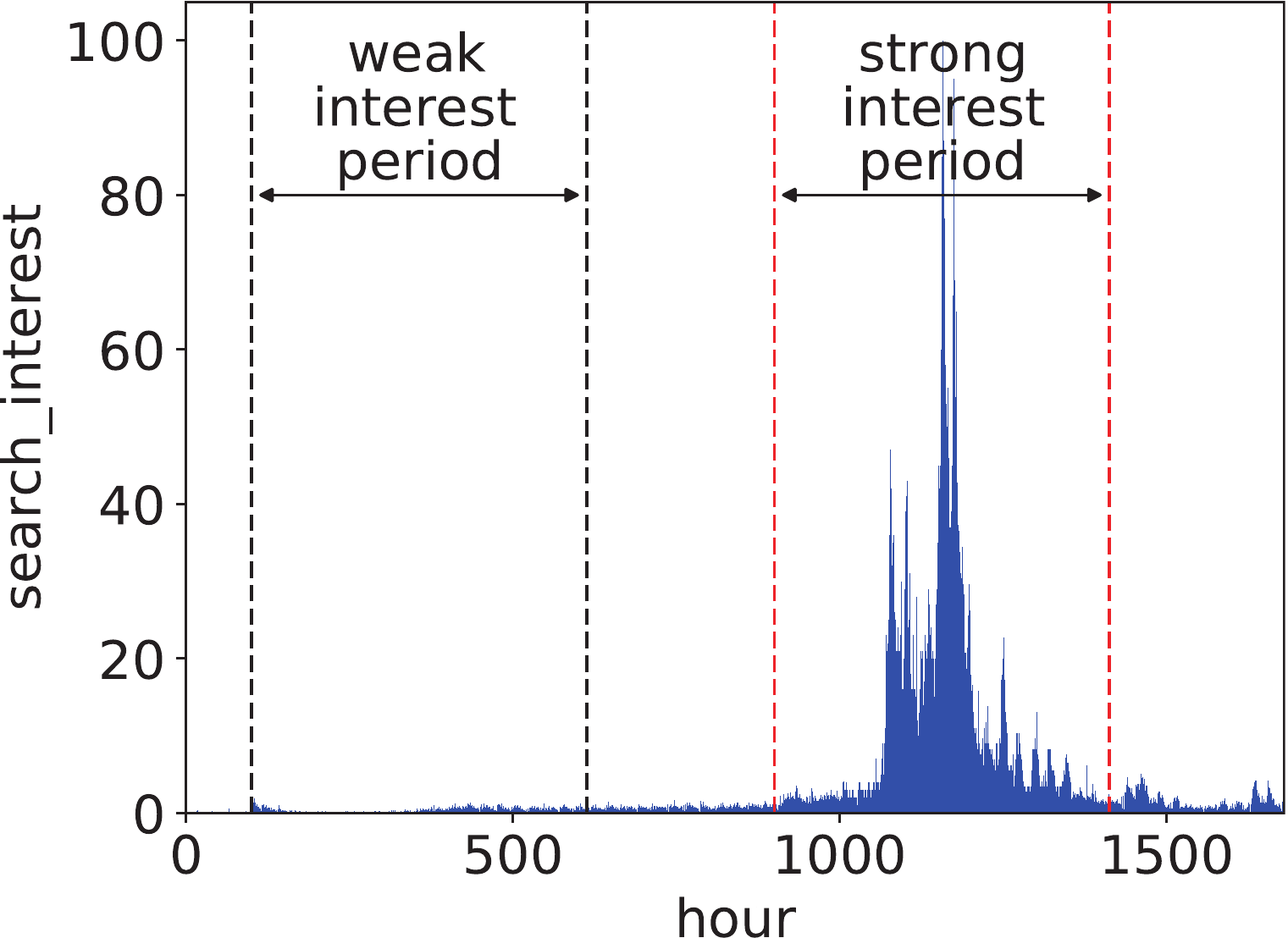} 
\includegraphics[width=0.48\linewidth]{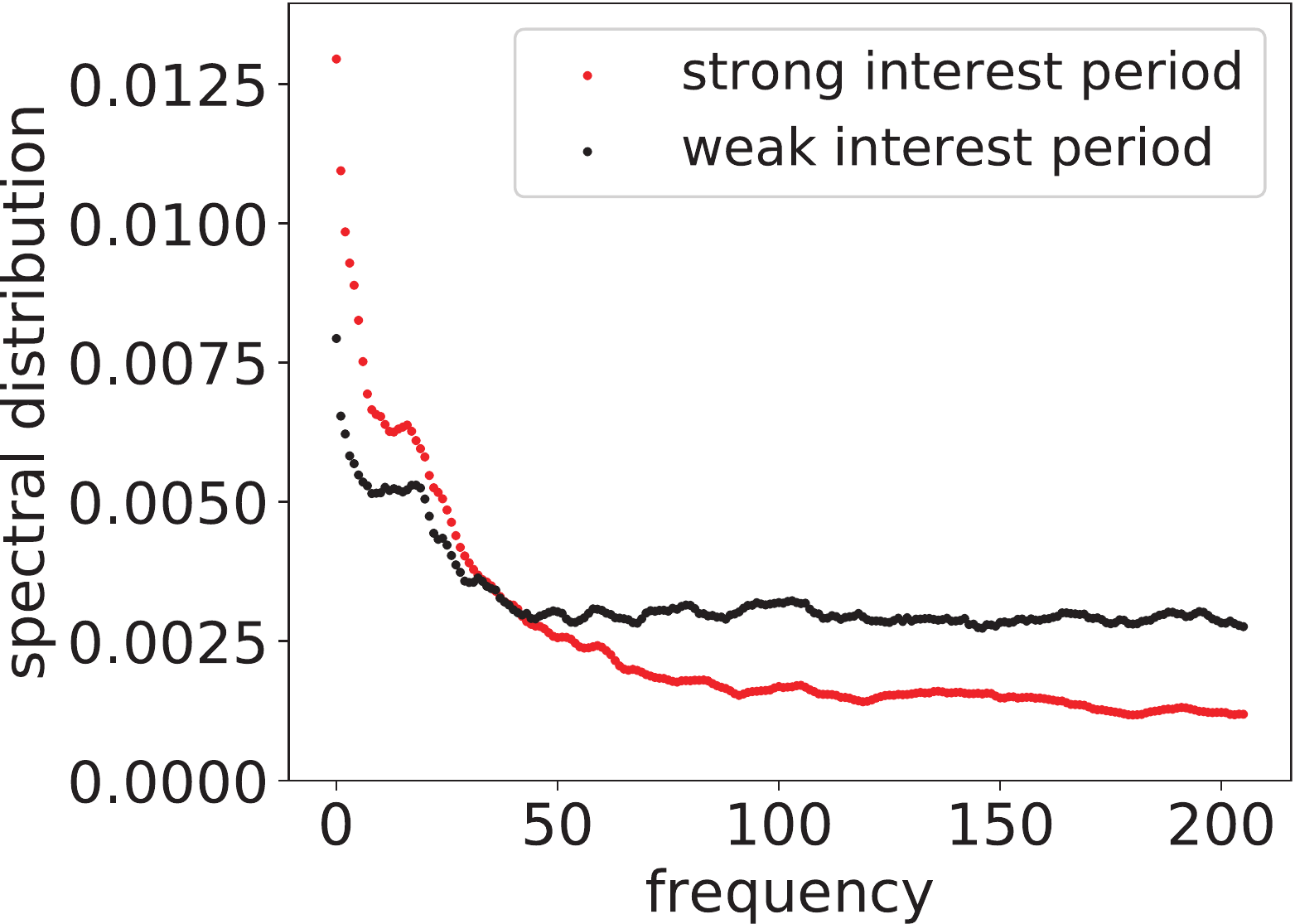}
\caption{Hourly transitions in interest in Japanese smartphone payment service ``7pay''(left) and moving average of the normalized spectral distribution with window size of $30$ (right)}
\label{7pay}
\end{center}
\end{figure}

The left panel of Fig.~\ref{7pay} shows the hourly transitions in the search interest for a Japanese smartphone payment service from May 1, 2019, to July 26, 2019; it also indicates the periods analyzed. 
The period indicated by black dotted lines had relatively few searches, while that indicated by red dotted lines had many searches.
The same color usage is adopted for other data sequences.
The right panel of Fig.~\ref{7pay} shows the normalized FFT (moving average with a window size of $50$) of the data in each period. 
This result shows that the period of more interest has stronger low-frequency modes.

\subsubsection{``Fukuoka Softbank Hawks'': a Japanese professional baseball team}
\begin{figure}[bt]
\begin{center}
\includegraphics[width=0.48\linewidth]{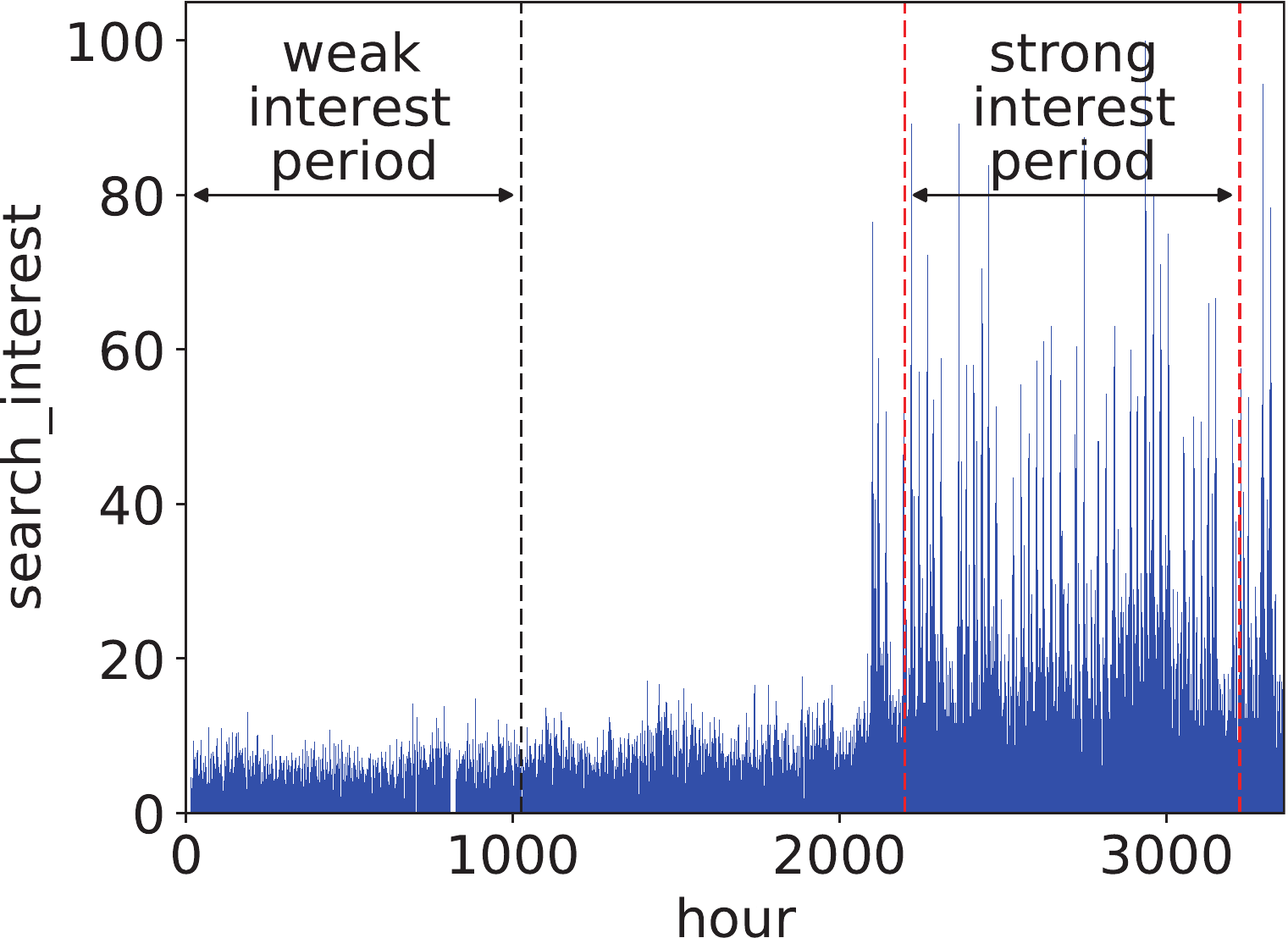}
\includegraphics[width=0.48\linewidth]{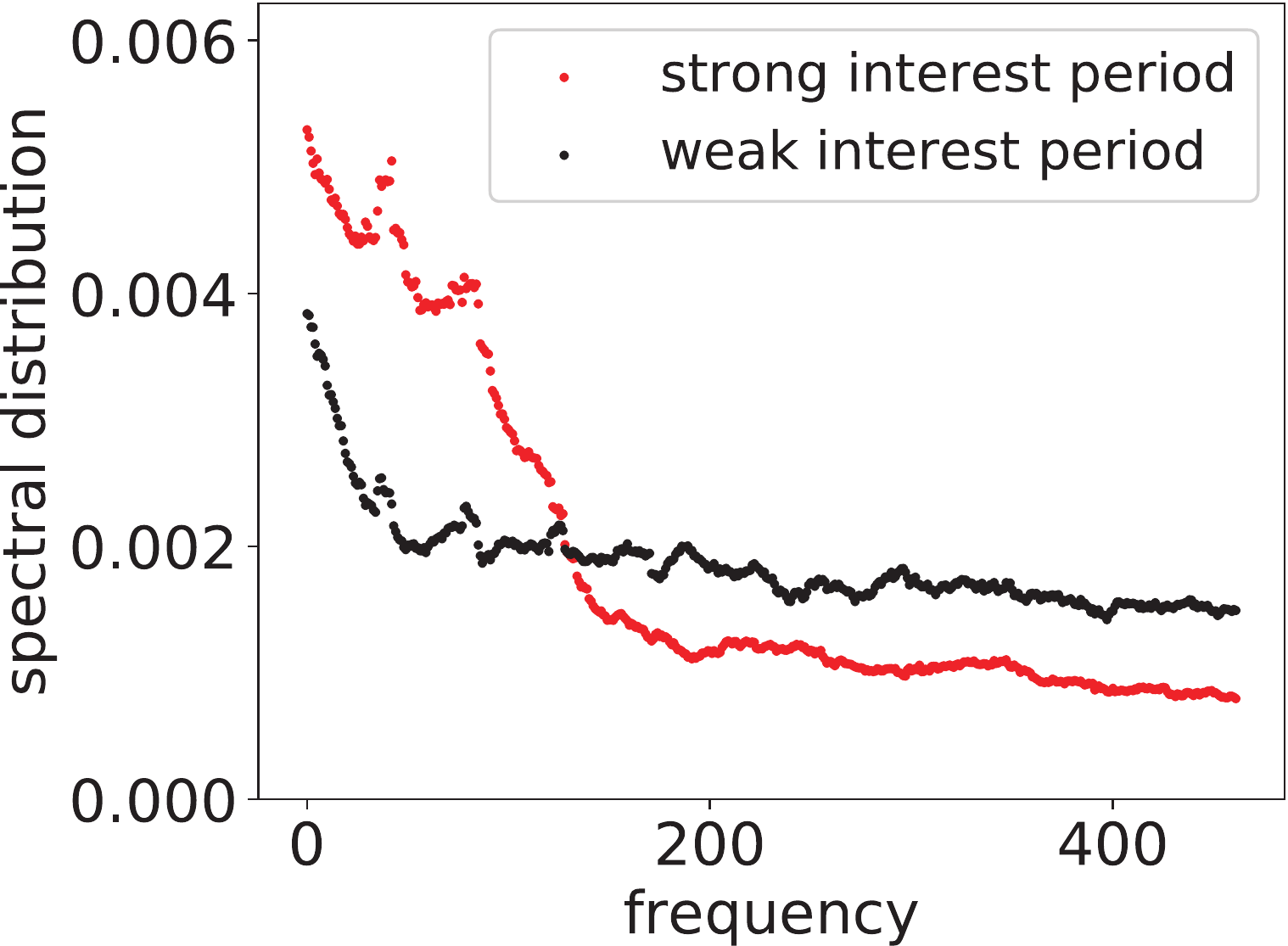}
\caption{Hourly transitions in the interest in Japanese professional baseball team ``Fukuoka Softbank Hawks''(left) and moving average of the normalized spectral distribution with window size of $50$ (right)}
\label{sh}
\end{center}
\end{figure}

The left panel of Fig.~\ref{sh} shows hourly transitions in the search interest for the Japanese professional baseball team ``Fukuoka Softbank Hawks'' from January 1, 2019, to May 21, 2019.
The right panel of Fig.~\ref{sh} shows the normalized FFT (moving average with a window size of $50$) of the data in each period. 
This result also shows that the period of more interest has stronger low-frequency modes.

\subsubsection{``fgo'': a smartphone game}
\begin{figure}[bt]
\begin{center}
\includegraphics[width=0.48\linewidth]{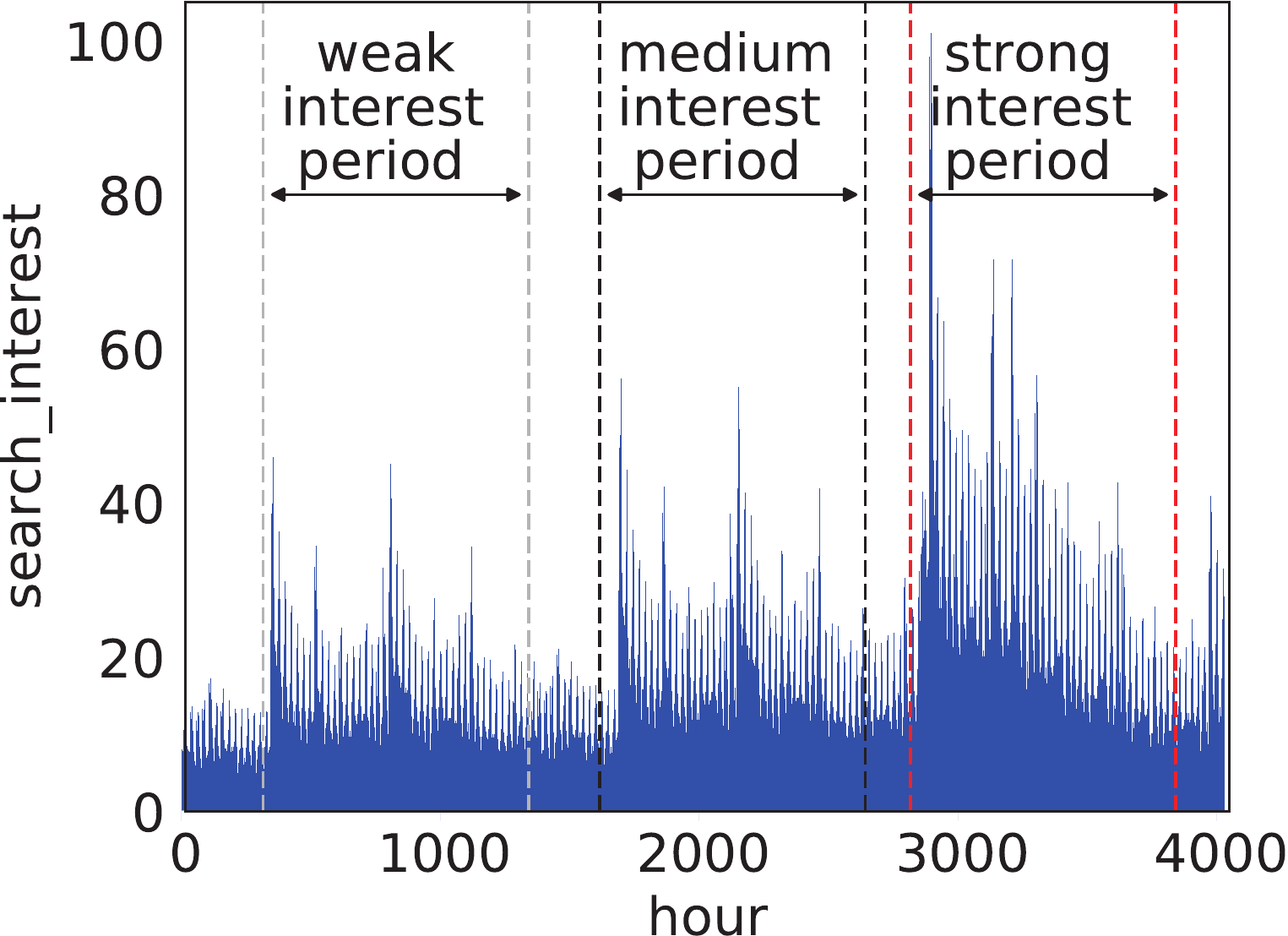}
\includegraphics[width=0.48\linewidth]{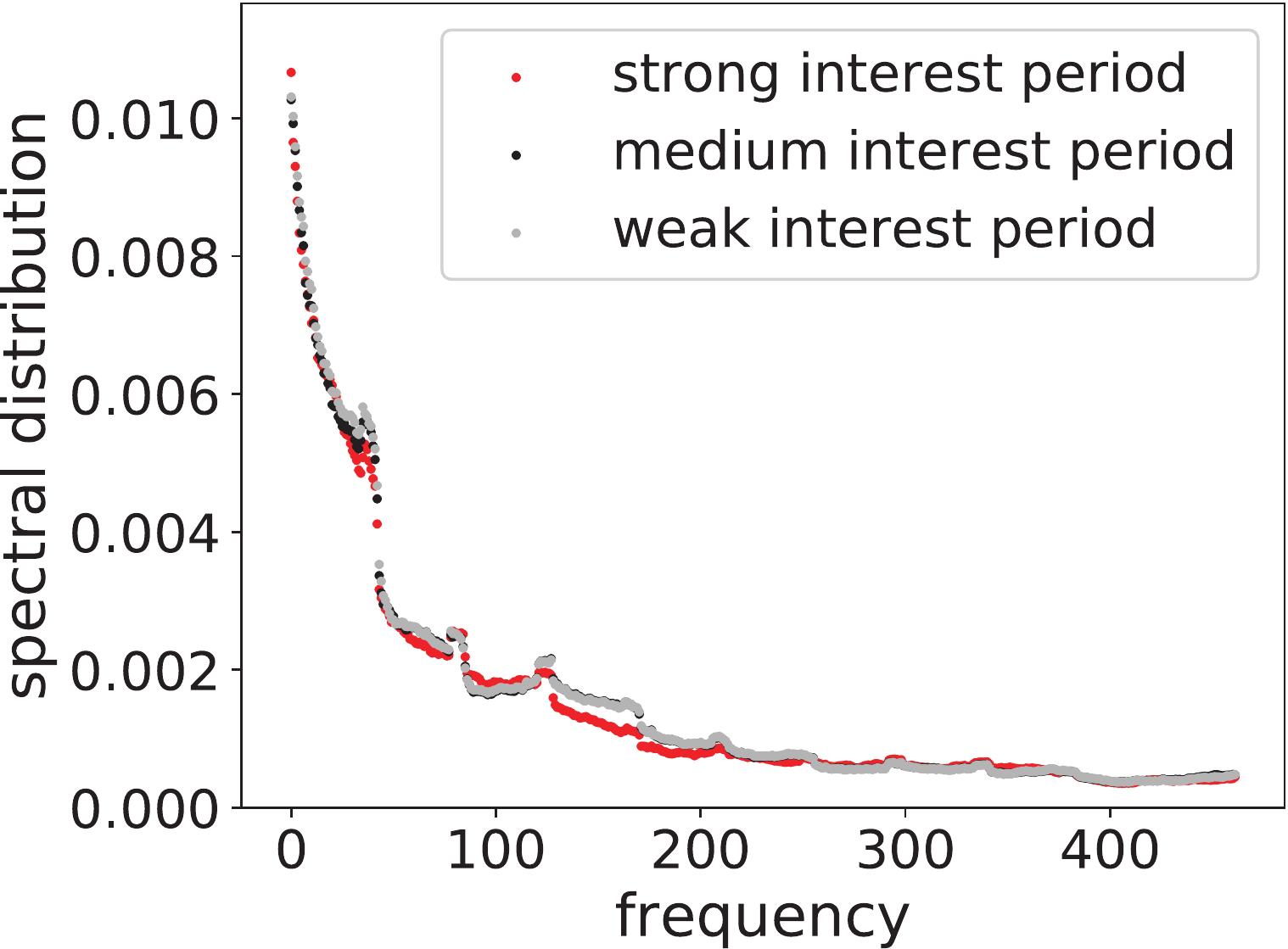}
\caption{Hourly changes in the interest in smartphone game ``fgo'' (left) and moving average of the normalized spectral distribution with window size of $50$ (right)}
\label{fgo}
\end{center}
\end{figure}

The left panel of Fig.~\ref{fgo} shows hourly changes in the search interest for the smartphone game ``fgo'' from June 1, 2019, to September 21, 2019.
The right panel of Fig.~\ref{fgo} shows the normalized FFT (moving average with a window size of $50$) of the data in each period. 
This result shows that all the periods exhibit similar spectrum distributions.  

\subsubsection{``shuukatsu'': job hunting}
\begin{figure}[bt]
\begin{center}
\includegraphics[width=0.48\linewidth]{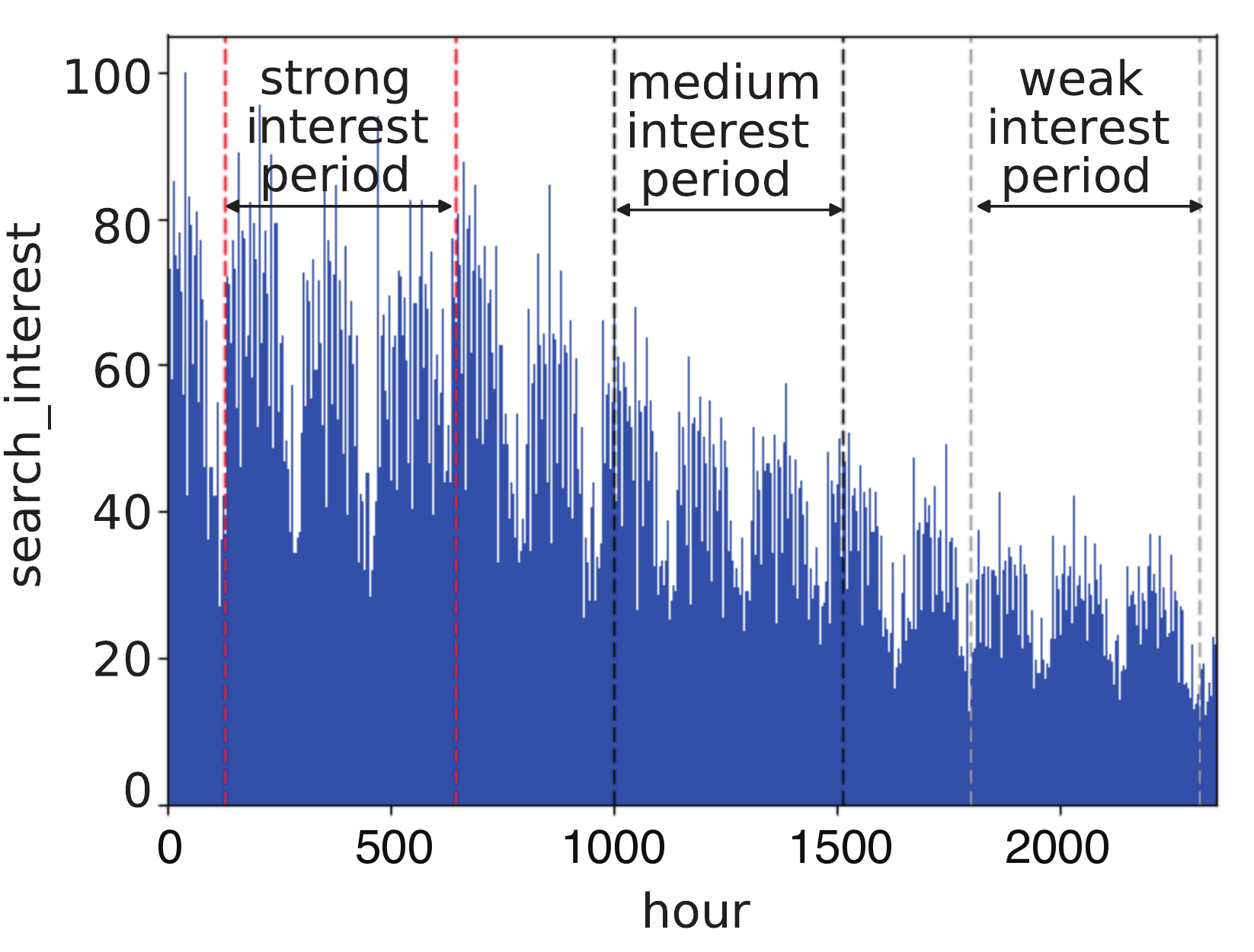}
\includegraphics[width=0.48\linewidth]{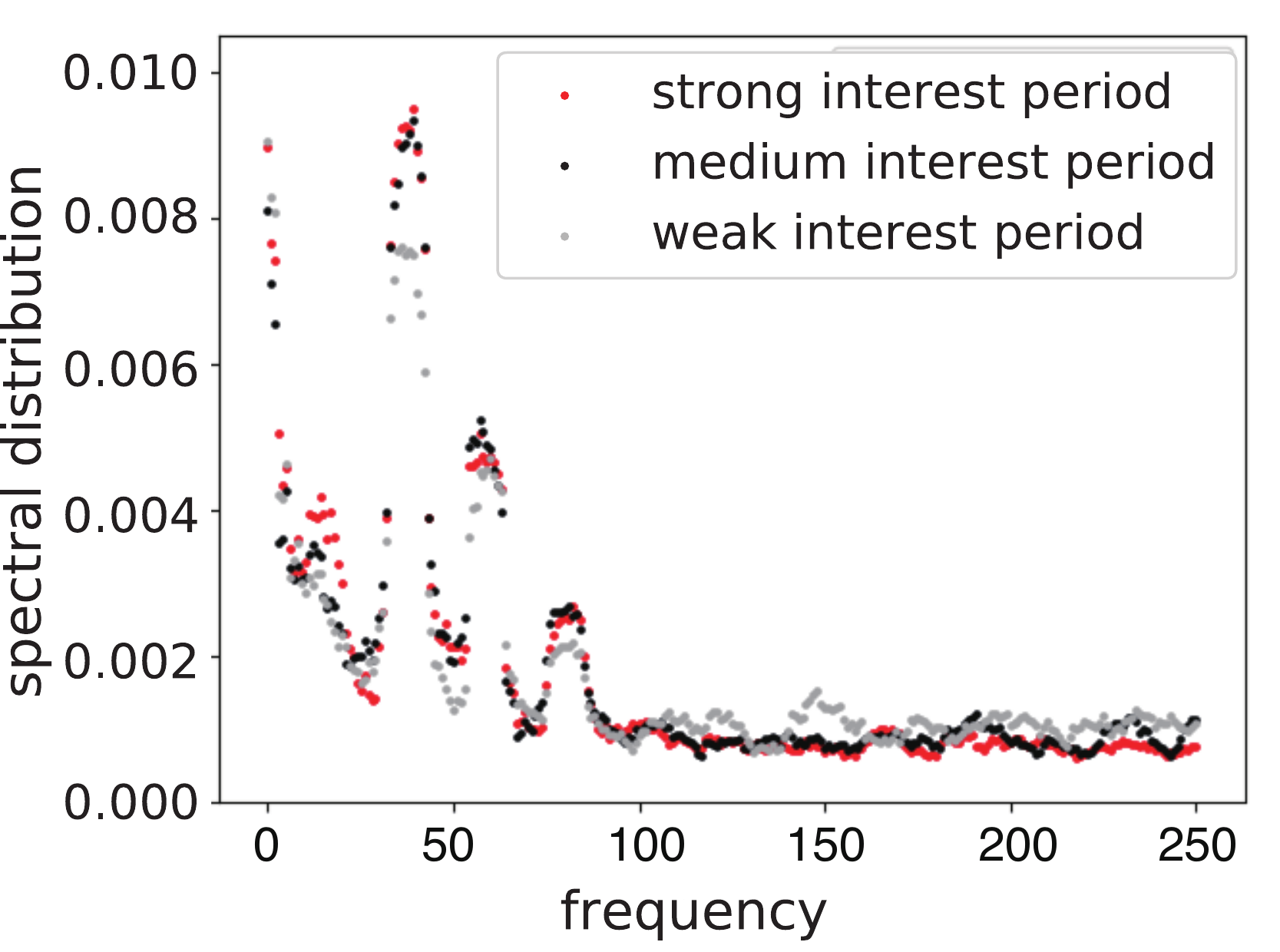}
\caption{Hourly changes in the interest in job hunting ``shuukatsu'' (left) and moving average of the normalized spectral distribution with window size of $10$ (right)}
\label{job}
\end{center}
\end{figure}

The left panel of Fig.~\ref{job} shows hourly transitions in the search interest for job hunting ``shuukatsu'' from May 7, 2019, to August 13, 2019.
The right panel of Fig.~\ref{job} shows the normalized FFT (moving average with a window size of $10$) of the data in each period. 
This result also shows that all the periods exhibit similar spectrum distributions.  

\subsubsection{``denshi kessai'': electronic payment}
\begin{figure}[bt]
\begin{center}
\includegraphics[width=0.48\linewidth]{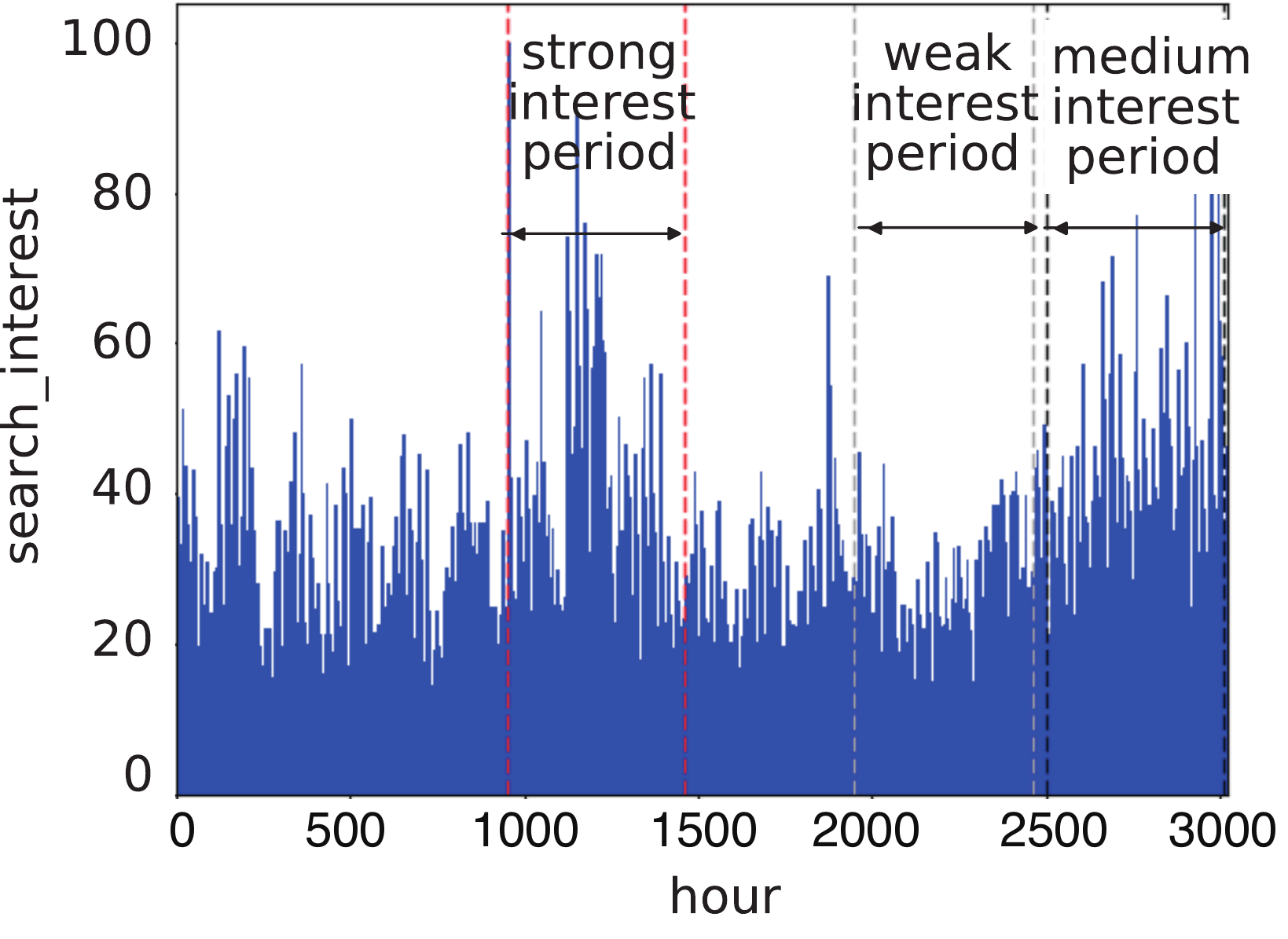}
\includegraphics[width=0.48\linewidth]{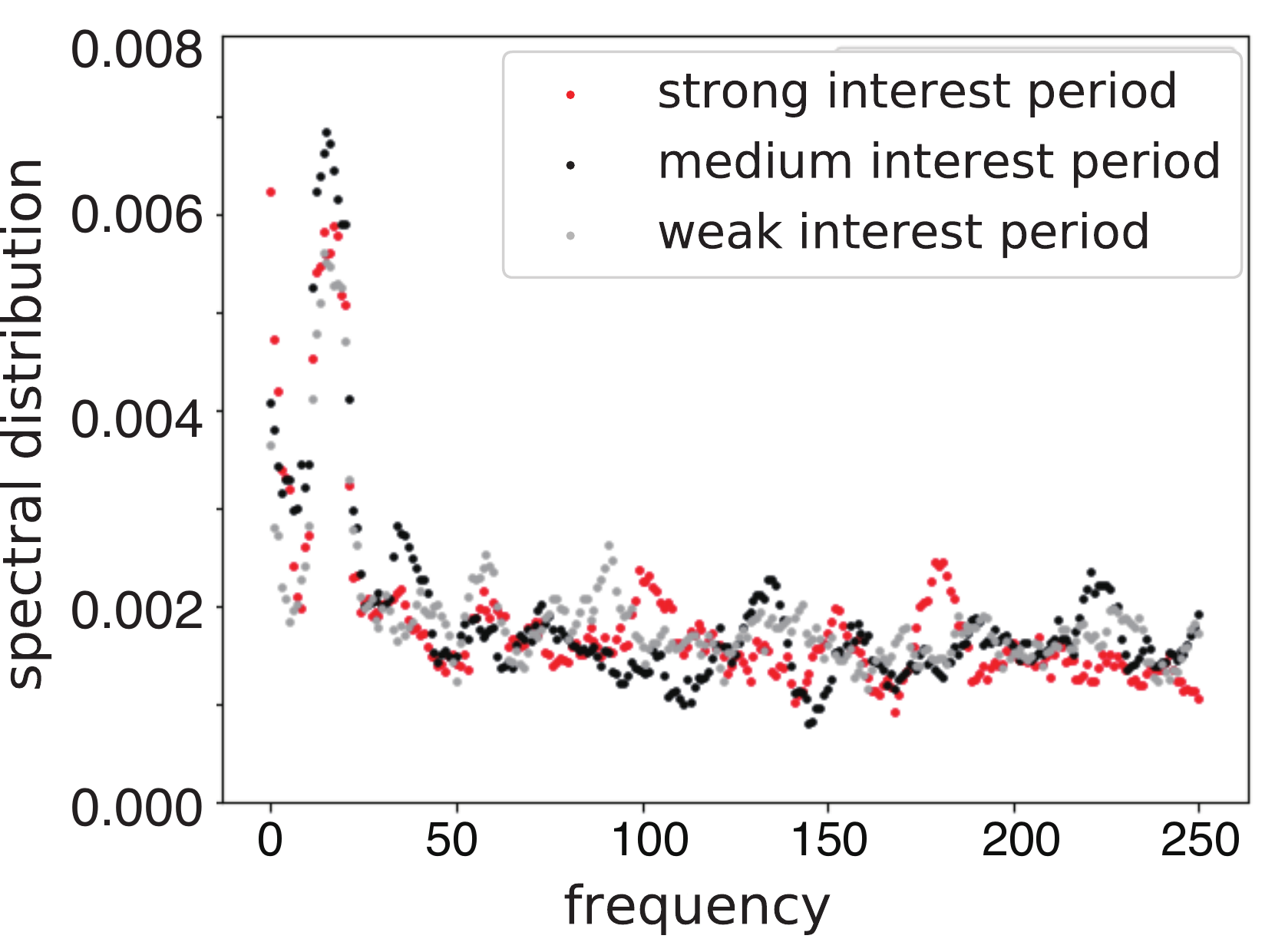}
\caption{Hourly changes in the interest in electronic payment ``denshi kessai'' (left) and moving average of the normalized spectral distribution with window size of $10$ (right)}
\label{denshi}
\end{center}
\end{figure}

The left panel of Fig.~\ref{denshi} shows hourly changes in the search interest for electronic payment ``denshi kessai'' from June 15, 2019, to September 18, 2019.
The right panel of Fig.~\ref{denshi} shows the normalized FFT (moving average with a window size of $10$) of the data in each period. 
This result also shows that all the periods exhibit similar spectrum distributions.  

Even if there is a difference in the strength of the search interest, it can be seen that the combination of periodic fluctuations and long-term trends without sudden excitement does not change the spectral distribution.
With all results gathered from Google Trends, the low-frequency mode of user dynamics increases when the topic that attracts attention emerges, but there is no change in normal fluctuations.
This supports the prediction that low-frequency modes become apparent when the structure of the OSN changes to activate online user dynamics, which is predicted from the oscillation model.

\section{Conclusion} 
\label{sec:conclusion}
Understanding user dynamics in OSNs is of significant interest to understand not only information networks but also real social activities. 
Many approaches have been introduced to investigate user dynamics. 
Most approaches used for describing the diffusion of information are based on the infectious disease model, such as the SIR model. 
While this approach expresses the state transition of the system in a macroscopic framework, it is not good at describing individual user dynamics. 
Other models use either the diffusion model or the continuous-time Markov chain, both of which are based on first-order differential equations with respect to time. 
Both models are good at describing the steady-state of the system but are inappropriate for describing explosive user dynamics. 

The oscillation model is based on the wave equation in OSNs, which is a second-order differential equation with respect to time. 
The remarkable features of the oscillation model include (1) it is based on a minimal model that describes the properties shared by a wide range of user dynamics, (2) it gives a unified interpretation of the conventional indices of node centrality, and it also yields the extended concept of node centrality, and (3) explosive user dynamics like flaming can be described. 
In addition, modeling by the wave equation is suitable for describing the situation in which the influence of interaction between users {\it propagates through the network at finite speeds}. 
However, since the oscillation model was, up to now, considered in purely theoretical terms, it was necessary to confirm whether the model describes real phenomena correctly or not.

We extracted from the oscillation model the prediction that the low-frequency oscillation mode would be dominant when the structure of OSNs changes to activate online user dynamics.
We confirmed the validity of the oscillation model by verifying this prediction with actual data.
Our tests included conducting spectral analyses of the log data of posts on electronic bulletin board sites and word search frequency from Google Trends. 
As the results of processing the data match the predictions, the oscillation model can appropriately describe the user dynamics.
The results confirm the prediction generated from the theory. 

If the oscillation model can be verified as suitable for describing online user dynamics, a technique for mitigating explosive user dynamics may be realized. 
The oscillation model yields two strategies. 
The first is a method that changes the link structure so that all eigenvalues of the Laplacian matrix representing the structure of the online social network are real numbers. This corresponds to operations such as delaying the time of entering a new post on the electronic bulletin board or delaying the arrival time of email. 
The other is to increase the damping strength of user dynamics. This corresponds to the operation of quickly attenuating the degree of attention to the topic currently drawing attention by transmitting new information different from the current topic of interest.

\section*{Acknowledgment}
This research was supported by Grant-in-Aid for Scientific Research (B) No.~19H04096 (2019--2021) and No.~20H04179 (2020--2022) from the Japan Society for the Promotion of Science (JSPS).


\end{document}